\documentclass[%
 aip,
 amsmath,amssymb,
 reprint,%
]{revtex4-1}
\usepackage{subfiles} 
\usepackage{graphicx}
\usepackage{dcolumn}
\usepackage{bm}

\usepackage[utf8]{inputenc}
\usepackage[T1]{fontenc}
\usepackage{mathptmx}
\usepackage{booktabs}
\usepackage{natbib}
\usepackage{amsmath}
\usepackage{bm}
\usepackage{mathrsfs}
\usepackage{array}
\usepackage{graphicx}
\usepackage{epstopdf, epsfig}
\usepackage{subcaption}
\usepackage{amsmath}

\usepackage[usenames, dvipsnames]{color}
\usepackage{color}
\usepackage{tikz}
\usepackage{xcolor}
\usepackage{soul}
\usepackage{ulem}
\usetikzlibrary{shapes,arrows}
\usepackage{booktabs}
\usepackage{varwidth}
\usepackage{rotating}
\usepackage{wrapfig}
\usepackage{dblfloatfix}
\usepackage[export]{adjustbox} 
\usepackage{tabularray}
\usepackage{float}
\usepackage{cancel}
\newsavebox\tmpbox
\newcommand\vol{\ooalign{\hfil$V$\hfil\cr\kern0.08em\raise2pt\rlap{--}\hfil\cr}} 
\newcolumntype{P}[1]{>{\centering\arraybackslash}p{#1}} 
\newcolumntype{M}[1]{>{\centering\arraybackslash}m{#1}} 
\newcommand{\av}[1]{\overline{#1}}
\newcommand{\mbx}{\mathbf{x}}
\newcommand{\mbr}{\mathbf{r}}

\newcommand{\rf}[1]{\eqref{#1}}

\newcommand{\lr}[1]{\left({#1}\right)}
\newcommand{\lrs}[1]{\left[{#1}\right]}
\newcommand{\pdr}[2]{\frac{\partial{#1}}{\partial {#2}}}

\newcommand{\xo}{( x^I)}

\newcommand{\ol}[1]{\overline{#1}}
\newcommand{\xt}{(x^{II})}
\newcommand{\tuk}{\overline{u}_k}
\newcommand{\tui}{\overline{u}_i}
\newcommand{\tuj}{\overline{u}_j}
\newcommand{\tT}{\overline{T}}

\newcommand{\buk}{\overline{u}_k}
\newcommand{\bui}{\overline{u}_i}
\newcommand{\buj}{\overline{u}_j}

\newcommand{\bT}{\overline{T}}

\newcommand{\eqn}{\begin{equation}}
\newcommand{\eqnn}{\end{equation}}
\newcommand{\spl}{\begin{split}}
\newcommand{\spll}{\end{split}}

\newcommand{\xtt}{(\bm{x+r})}
\newcommand{\x}{( \bm{x})}

\begin{document}

\preprint{AIP/123-QED}

\title{\textcolor{black}{Scale-space energy transport in anisotropic inhomogeneous buoyancy-driven turbulent flows}}

\author{Endale H. Kirubel}
\email{kirubel.endale.haile@gmail.com }
\affiliation{
Department of Aerospace Engineering, Indian Institute of Technology Madras,
Chennai-600036, India
}

\author{P. Aswin}
\email{ae21d030@smail.iitm.ac.in}
\affiliation{
Department of Aerospace Engineering, Indian Institute of Technology Madras,
Chennai-600036, India
}

\author{A. Sameen}
 \email{sameen@ae.iitm.ac.in}
\affiliation{
Department of Aerospace Engineering, Indian Institute of Technology Madras,
Chennai-600036, India
}
 \affiliation{Geophysical Flows Lab, Indian Institute of Technology Madras, Chennai-600036,
India}

\date{\today} 

\begin{abstract}
In this article, we introduce a new mathematical framework that can describe the budget of turbulence kinetic energy and heat transfer in both physical space and scale space of turbulence. We derived two exact transport equations for inhomogeneous and anisotropic buoyancy-driven incompressible turbulent flow. The two equations are the turbulent kinetic energy transport equation and the turbulent heat flux density transport equation. These equations are derived from the governing Navier-Stokes equation and two-point correlations. Analogous to the energy spectrum and heat flux spectrum for homogeneous isotropic turbulence, energy density, and heat flux density introduced in this paper describe the energy and heat flux corresponding in various scales and the transport across physical and scale space for inhomogeneous anisotropic turbulent flows. \textcolor{black}{The proposed equations are used to post-process a turbulent Rayleigh-B\'enard convection flow dataset. Specifically, production, dissipation, interscale transport, and buoyancy terms are evaluated, and the contribution from inhomogeneity is quantified. }
\end{abstract}

\maketitle

\section{Introduction} \label{sec:intro}
Turbulence has been known as a unique characteristic of a flow for nearly five centuries. \textcolor{black}{ It is characterized by chaotic-like fluid motions and distinct vortical structures. Vortices of different length scales are present in a typical turbulent flow as illustrated by Da Vinci\cite{Da_vinci}.} \textcolor{black}{ In his pioneering experiment\cite{reynolds1883xxix}, Reynolds demonstrated the chaotic nature of turbulent flow and examined the transition of the flow from laminar to turbulent in a pipe. Though this experiment dates back to 1883, the physics of turbulence is not yet completely explained by theories that currently exist.} At the beginning of the last century, attempts were made to have a mathematical framework developed to capture its nature.  Taylor introduced the statistical approach \cite{Taylor_1935}, and Kolmogorov presented some important and widely used turbulence theories \cite{kolmogorov1941dissipation}. His theory, popularly known as K41, enabled the predictions of the energy spectrum for a homogeneous and isotropic turbulent flow. The complexity in turbulence flow analysis is mostly due to the perplexing energy transfer between different eddy sizes(scales), and interactions between different flow regions \citep{Arun_Sameen_Srinivasan_Girimaji_2021}.

Solving the Navier-Stokes equation directly is complex and computationally costly for many practical applications. Often, RANS/LES models are used to solve turbulence problems where the effect of small-scale turbulence is modeled. These models used the turbulence dissipation rate as a surrogate for the energy transfer across scales, and Kolmogorov's 4/5 law \cite{kolmogorov1962refinement} links the interscale energy transfer with the dissipation at a given physical space point.  For a statistically stationary flow, the dissipation rate is equivalent to the production at large scales, and it is the same as the rate of energy transferred at each scale through the cascade. We can apply the Fourier transform and observe the energy distribution among scales in the wavenumber space, the energy spectrum. There are many who have contributed significantly to the understanding of homogenous isotropic turbulence \cite{obukhov1941distribution, obukhov1941spectral, richardson1926atmospheric, richardson1948note, kolmogorov1941local, kolmogorov1962refinement, kraichnan1971inertial, kraichnan1975remarks, hinze1975turbulence, frisch1995turbulence, benzi2015homogeneous}
  The validity of the Richardson-Kolmogorov cascade scheme is observed in small scales at high Reynolds number flows, even when the flows are inhomogeneous \cite{gage1986theoretical, gargett1984local, grant1962turbulence, champagne1977flux, browand1983mixing, gibson1963spectra}.   Anisotropy and inhomogeneity, however, will affect the energy distribution and the energy
transfer between scales, which is investigated in this paper.  Since the Fourier transform is not applicable along inhomogeneous directions, wave number space for scale space description is ill-suited. 



\textcolor{black}{Almost all turbulent flows in nature or engineering applications are inhomogeneous and anisotropic in nature. We need to develop a framework that can accurately describe the energy/heat transfer among scales for these flows. Different researchers have made progress in formulating the scale space equivalent of the energy spectrum for inhomogeneous turbulence. One such candidate is the second-order velocity structure function,  which is considered the scale energy. The transport equation of second-order structure function in scale space was investigated by many researchers \cite{KH_1938, Naot_1973, danaila2001turbulent, hill2002exact, davidson2005identifying}. 
K\'arm\'an and Howarth \citep{KH_1938} (KH) were the first to analyze a two-point correlation equation. The analysis showed the decay of isotropic turbulence and the evolution of the turbulent length scale. Although the KH equation was first derived for homogeneous isotropic turbulence, a further step was taken by Naot et al. \cite{Naot_1973}, where they extended the isotropic form of two-point correlation into a quasi-isotropic model. Danaila\cite{danaila2001turbulent} further modified the KH equation and accounted for inhomogeneity in a channel flow. A generalized Kármán–Howarth–Monin (KHM) equation was developed later by  \citet{hill2002exact} for inhomogeneous anisotropic turbulence. The generalized KHM equation provides great insight into the transport of energy in the physical and scale space. However, it is not the energy density equivalent of the energy density spectrum in the wave number space we need for inhomogeneous turbulence.  \citet{davidson2005identifying} proposed that the derivative of structure-function in scale space can describe the energy density in scale space.}

\textcolor{black}{Marati et al.\cite{Marati2004} used the generalized Kolmogorov equation from \citet{hill2002exact} to analyze the dynamics of wall-bounded turbulence using a low-Reynolds-number direct numerical simulation (DNS) of turbulent channel flow. They identified the existence of a coupled energy transfer occurring simultaneously in physical and scale space, which is anticipated by other researchers\cite{JIMENEZ1999252, townsend1976structure}. \citet{Cimarelli2012} studied the scale space energy transport using filtered velocities of a DNS simulation of turbulent channel flow. Their analysis demonstrated the importance of reverse energy transfer and the importance of modeling those interactions for accurately capturing turbulence in a large eddy simulation. Cimarelli et al.\cite{Cimarelli2013} analyzed the scale-energy interactions in three-dimensional scale space to identify the geometry of fluxes and sources. They observed that the fluxes follow a spiral-like path, where inverse energy transfer plays a crucial role. \citet{Mollicone2018} used the generalized Kolmogorov equation to study the scale space energy transport in turbulent shear layers and separation bubbles. They used the generalized Kolmogorov equation and its Lagrangian view to analyse the statistical evolution of the coherent structures in a shear layer. All the above articles use the structure function-based definition of scale energy and its extension to inhomogeneous anisotropic turbulence.}


On the other hand,  \citet{osti_79052} derived a transport equation for two-point velocity correlations. 
Further, \citet{hamba_2015} argued that when the separation distance $|\bm{r}|$ between two points is quite large, the second-order structure function simplifies to the sum of twice the turbulent kinetic energy values at these far-apart points. The two kinetic energy values can vary significantly if $\bm{r}$ does not lie along a homogeneous direction. Therefore, establishing a direct link between the transport equation for the structure function and the equation for turbulent kinetic energy at a single point is challenging. He then proposed that the two-point correlation as a better candidate for scale energy, and its derivative in scale space represents the energy density.  \citet{Arun_Sameen_Srinivasan_Girimaji_2021} derived a scale-space energy density transport equation for inhomogeneous compressible turbulent flows. These works show a good effort and progress in understanding the scale-space dynamics of turbulent kinetic energy. A velocity-temperature two-point correlation can be a powerful tool to analyze heat transport across scales. Progress in scale-space dynamics of turbulent heat transfer and internal energy is not attempted to the best of our knowledge.

In buoyancy-driven turbulence, buoyancy forces play a significant role in how energy is distributed among scales \cite{kadanoff2001turbulent, Taylor_1935, batchelor1959small, corrsin1951spectrum, ahlers2009heat, lohse2010small, chilla2012new, niemela2000turbulent, niemela2003rayleigh, niemela2010turbulent, amati2005turbulent, jiang2018controlling, stevens2018turbulent, oboukhov1949structure}.
A discussion on homogeneous turbulent flow with buoyancy taking an active role can be found in some of the recent books \cite{Favier2020, canuto2009, verma2018physics}. \textcolor{black}{The Rayleigh-B\'enard convection (RBC) flow, which describes the convective flow of fluid between two parallel plates heated from below and cooled from above, is one of the most studied canonical cases for buoyancy-driven flows. The various aspects of RBC flows have been studied over the years \cite{benard1900etude, rayleigh1916lix, kadanoff2001turbulent, ahlers2009turbulent, VERZICCO_CAMUSSI_2003, Weiss2023, heat_transport,Zhou2024}.} The scale energy and scale heat flux movement in eddy scale space in an inhomogeneous direction in a buoyancy-driven flow is not examined in the literature. \textcolor{black}{The scale energy density and scale heat flux density, which are derived in this paper, are evaluated for RBC to demonstrate the inferences that can be made from this analysis.}

This paper aims to introduce a new mathematical framework (governing transport equations) for kinetic energy and heat flux for a buoyancy-driven inhomogeneous turbulent flow. These new transport equations are derived from the Navier-Stokes equation and incompressibility condition without any other assumptions related to anisotropy or inhomogeneity.  
Various factors contributing to the overall balance of energy and heat transfer in physical space as well as in the scale space of turbulence for incompressible, anisotropic,  inhomogeneous flows can be deduced.
 The equations have the potential to advance our understanding of energy cascades and the contribution of each scale in energy and heat transport, thus improving turbulence modeling. 
The contents of the paper are in the following sequence. The mass, momentum, energy equation, and scale space transformation used are presented in Sec \ref{sec:gove}. The scale energy equation is derived in Sec \ref{sec:energy} and the scale heat flux equation in Sec \ref{sec:heat}. Further, the energy density and heat flux density in scale space describing the transport from eddy to eddy is discussed in Sec \ref{density}. \textcolor{black}{Finally, the analysis of the turbulent RBC dataset using these equations is discussed in Sec \ref{sec:RBC}.}

\section{Governing equation and scale transformation}
\label{sec:gove}
 The governing equations for incompressible buoyancy-driven flows (the conservation equations of mass, momentum, and energy), under Boussinesq approximation, can be expressed using index notation as,
\begin{equation}
    \pdr{u_k}{x_k} = 0 ~ , \label{uu1}
\end{equation}
\begin{equation}
    \pdr{u_i}{t} + u_k \pdr{u_i}{x_k} = -\frac{1}{\rho}\pdr{p}{x_k}\delta_{ik} + \nu \pdr{^2u_i}{x_k^2} + g_i \beta \lr{T-T_0}~, \text{and}
    \label{uu2}
\end{equation}
\eqn{\pdr{T}{t} + u_k\pdr{T}{x_k} =\alpha \pdr{}{x_k}\pdr{T}{x_k},
\label{t5}}
\eqnn

 where $u_i$ is the component of velocity vector,  $T$ is temperature, $p$ is  pressure,  ${\rho}$ is the reference density, \textcolor{black}{$\delta_{ik}$ is the Kronecker delta, $T_0$ is the reference temperature,} $g_i$  is the gravitational acceleration with the form $(0, -g \hat{e}_y, 0)$, and $\hat{e}_y$ is the unit vector in the y direction. Fluid properties kinematic viscosity is $\nu$, and thermal diffusivity is $\alpha$.

We use Reynold's averaging technique to decompose the instantaneous quantities into mean and fluctuating quantities  as
\begin{align}
& u_i=\av{u_i} + u_i',  \\
& p = \av{p} + p' , \\
& T = \av{T} + T '.
\end{align}
The fluctuating quantities from Reynolds averaging are represented by a single prime $(p', u', T')$, whereas mean quantities are accompanied by the over-bar $(\av{u},~ \av{p}, ~ \av{T})$. Averaging Eqn \rf{uu1}, Eqn \rf{uu2}, and Eqn \rf{t5}, will give the mean governing equations as: 
\eqn
{\pdr{\av{u_k}}{x_k} =0 ,
\label{uu8} }
\eqnn
\eqn{
     \pdr{\bui}{t} + \buk \pdr{\bui}{x_k} + \textcolor{black}{\pdr{R_{ik}}{x_k}} = -\frac{1}{\rho}\pdr{\ol{p}}{x_k}\delta_{ik} + \nu \pdr{^2\ol{u_i}}{x_k^2} + g_i \beta (\ol{T}-T_0), \label{uu10}}
\eqnn
\begin{align}
   \pdr{\tT}{t} + \tuk \pdr{\tT}{x_k} & + \pdr{E_k}{x_k}= \alpha \pdr{}{x_k}\pdr{\tT}{x_k},
\label{t16} 
\end{align} 
where $E_k= \ol{ u_k' T'}$, and $R_{ik}= \ol{ u'_k u'_i}$.

By subtracting the mean equations (Eqn\rf{uu10}, and Eqn\rf{t16}) from the instantaneous equations (Eqn \rf{uu2}, and Eqn \rf{t5}), we obtain the governing equations for the fluctuating quantities as:
\begin{align}
    &\pdr{u_i'}{t} + u_k \pdr{u_i'}{x_k} + u_k' \pdr{\tui}{x_k} - \pdr{R_{ik}}{x_k} =  -\frac{1}{\rho}\pdr{{p'}}{x_k}\delta_{ik} + \nu \pdr{^2{u'_i}}{x_k^2} + g_i \beta T' ,
\label{uu12}
\end{align}
\begin{align}
    & \pdr{T'}{t} + u_k \pdr{T'}{x_k} + u_k' \pdr{\tT}{x_k} - \pdr{E_k}{x_k} =  \alpha \pdr{}{x_k}\pdr{T'}{x_k},
    \label{t20}
\end{align}

Equation \rf{uu12} can be rewritten by changing the indice $i$ with $j$ as,
\begin{align}
    \pdr{u_j'}{t} + & u_k \pdr{u_j'}{x_k} + u_k' \pdr{\tuj}{x_k} - \pdr{R_{jk}}{x_k} =  -\frac{1}{\rho}\pdr{{p'}}{x_k}\delta_{jk} + \nu \pdr{^2{u'_j}}{x_k^2} + g_j \beta T' .
\label{uu13}
\end{align}

\begin{figure}
    \centering
    \includegraphics[width=0.95\linewidth]{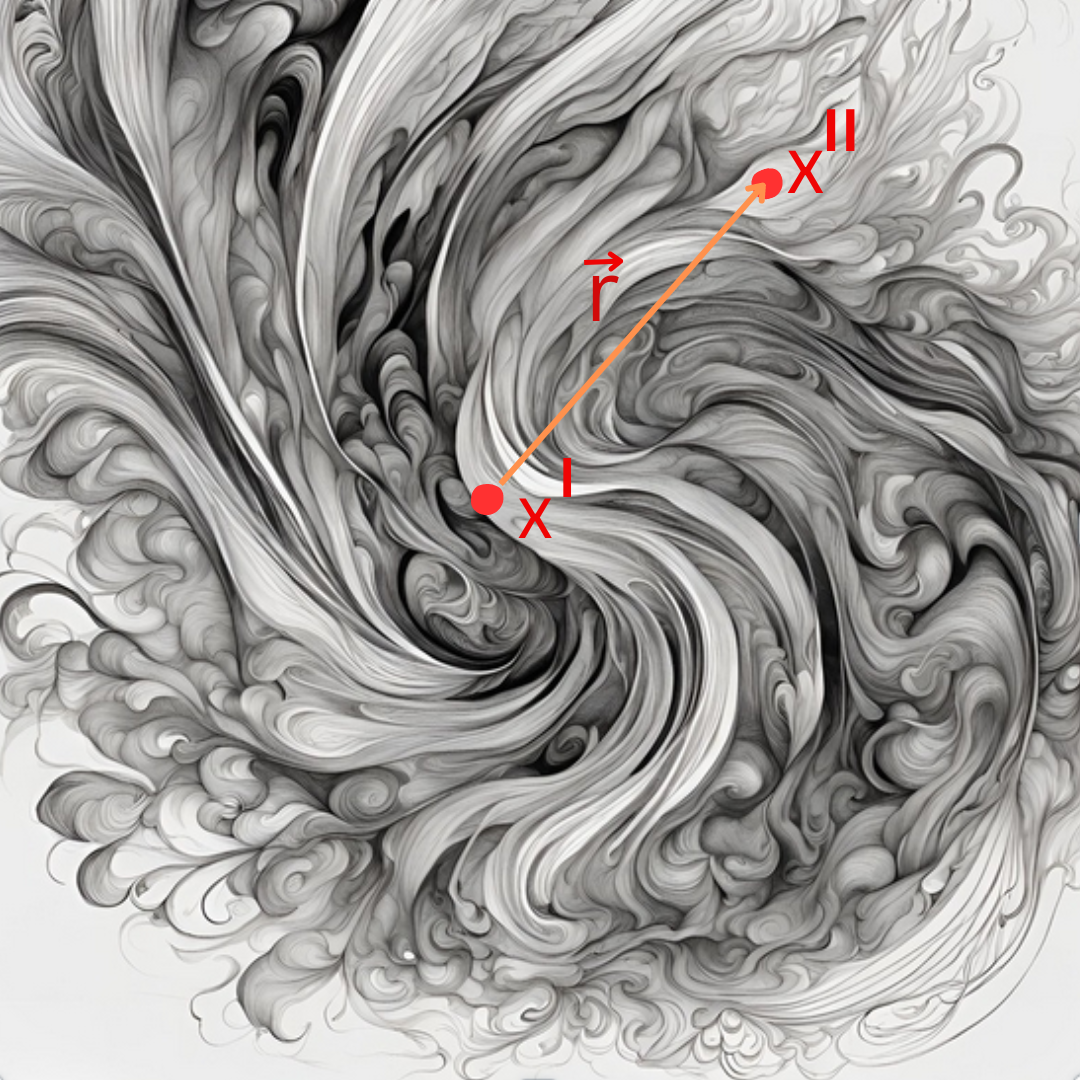}
    \caption{
    Schematic of an inhomogeneous turbulent flow indicating the coexistence of eddies of various sizes. From the two-point correlation analysis, we aim to determine the energy/heat transfer among different scales.
    }
    \label{fig:inhomo}
\end{figure}

Equation \rf{uu12}, \rf{t20}, and \rf{uu13}, along with the two-point correlations, are the equations from which an exact transport equation for energy and heat transfer in the scale space and physical space will be derived.
The two points are represented by variables $x^I$ and $x^{II}$, and they are considered independent, i.e., no relative motion \citep{hill1997applicability}:
\eqn{
\frac{\partial f(x^{II})}{\partial x_i^I} = \frac{\partial f(x^I)}{\partial x_i^{II}}=0 } ~,
\label{cross}
\eqnn 
for any function $f$.
The subscript $i,~ j,$ and $k$ notation is reserved to indicate directions only. The notations $\partial/\partial x_i^I$, and $\partial /\partial x_i^{II}$ will be used to represent the derivatives with respect to $x_i ,~ (=x,y,z)$ at points $x^I$  and  $x^{II}$, respectively.

\subsubsection*{Scale Transformation}

It is to be noted that velocity, pressure, and temperature obtained from numerical simulation of Navier-Stokes and energy equation are functions of physical space and time.
To obtain the transport equations, which can capture the transfer/transport of kinetic energy and heat flux in the physical and scale space, we apply a coordinate transformation \cite{osti_79052, hill2002exact, hamba_2015, Arun_Sameen_Srinivasan_Girimaji_2021}. The transformed equations are written in terms of physical space, $\bm{x}$, and in scale space $\bm{r}$. 
\textcolor{black}{Consider two points in space \textbf{$x^I$} and \textbf{$x^{II}$} separated by distances $\bm{r}$ as shown in schematic in Fig \ref{fig:inhomo}. The first point \textbf{$x^I$} is referred to as the center coordinates, and the separation distance $\bm{r}$ between the first and second point \textbf{$x^{II}$} is referred to as the relative coordinate. }
 The coordinates transform as, 
\begin{align}
   &  x^I= \bm{x} , \label{trans}\\
   & x^{II}= x^I+ \bm{r} = \bm{x}+\bm{r}, \\ 
   & \bm{r}=x^{II} - x^{I} ,
\end{align}
and the derivative operators transform as,
\begin{align}
    & \frac{\partial ()}{\partial x_k^{I}}= \frac{\partial ()}{\partial x_k} - \frac{\partial ()}{\partial r_k},
    \\
    & \frac{\partial ()}{\partial x_k^{II}}= \frac{\partial ()}{\partial r_k} ,
    \\
    & \pdr{}{x^I_k}\pdr{}{x^I_k}= \pdr{^2}{x_k^2} - 2\pdr{}{x_k}\pdr{}{r_k} + \pdr{^2}{r_k^2} ,
     \\
    & \pdr{}{x_k ^{II}}\pdr{}{x_k ^{II}}= \pdr{^2}{r_k^2} .
    \label{transf}
\end{align}
We applied these transformations in the subsequent sections while deriving the scale space transport equations for the kinetic energy and heat flux.

\section{Scale energy transport equation} \label{sec:energy}
\label{sec-Qii}

Two-point velocity correlation for an incompressible flow is defined as, 
\begin{align}
Q_{ij}(x^I, x^{II} )=Q_{ij}(\bm{x},\bm{x}+\bm{r})= \ol{u_i'\xo u_j'\xt} = \ol{q_{ij}(x^I, x^{II} )}.
\end{align}
This two-point correlation is a tensor quantity. If we take the trace of the tensor only (identical indices), we will get $Q_{ii}$, which plays the role of scale kinetic energy. In a one-point limit where $|\bm{r}|$ goes to zero, the two points $x^I ~\text{and} ~x^{II}$ are the same, and $Q_{ii}=\av{u'_i(\bm{x})u'_i(\bm{x})} = 2 KE$ is twice the total turbulence kinetic energy at that physical point $x$. On the other hand, if we take $|\bm{r}|$ to be very large ($|\bm{r_L}|$), $Q_{ii} $ would have a zero value because the two points separated by a very large distance $|\bm{r_L}|$ would be uncorrelated. From this, we can conclude that for any $\bm{r}$, where $0<|\bm{r}|<|\bm{r_L}|$, $Q_{ii}(\bm{x},\bm{r})$ is a part of turbulence kinetic energy with scale greater than $|\bm{r}|$, and we refer $Q_{ii}$ as scale kinetic energy.
\textcolor{black}{Turbulent flow consist of large number of scales from the largest integral length scale to the smallest Kolmogorov scale. Here, the term $Q_{ii}$ which is called the scale energy, gives the energy contained in all the scales from the selected one to the largest scale of the flow.}


The evolution of unaveraged two-point correlation ${q_{ij}}$, is then of the form,
\begin{align}
   \pdr{q_{ij}}{t} +  \tuk \pdr{q_{ij}}{x_k}
  &= {u_i'\xo} \lr{ \pdr{u_j' \xt}{t} + \tuk \xt \pdr{u_j'\xt}{x^{II}_k}}
     \nonumber \\
     & +{ u_j'\xt} \lr{ \pdr{u_i' \xo}{t} + \tuk\xo\pdr{u_i'\xo}{x^I_k}}.
    \label{uu016}
\end{align}

Comparing Eqn\rf{uu016} with Eqn\rf{uu12}, and Eqn\rf{uu13} a two-point transport equation can be derived by multiplying Eqn\rf{uu12} by $ u_j'\xt$ , and Eqn\rf{uu13} by ${u_i'\xo}$.

\begin{align}
     & u_j'\xt\pdr{u_i'\xo}{t} + u_j'\xt \pdr{\buk \xo u_i'\xo}{x^I_k} +  u_j'\xt \pdr{u'_k\xo u_i'\xo}{x^I_k} \nonumber \\
    &  +u_j'\xt  \pdr{u_k'\xo \bui \xo}{x^I_k} - u_j'\xt \pdr{R_{ik}}{x^I_k} = - \frac{1}{\rho}u_j'\xt\pdr{p'\xo}{x^I_k}\delta_{ik}
     \nonumber \\ 
    & + \nu u_j'\xt \pdr{}{x^I_k}\pdr{{u_i'\xo}}{x^I_k} + g_i \beta u_j'\xt{T'\xo} ~.
     \label{urbc13}
 \end{align}
\begin{align}
   & u_i'\xo\pdr{u_j'\xt}{t} + u_i'\xo \pdr{\buk \xt u_j'\xt}{x^{II}_k} +  u_i'\xo \pdr{u'_k\xt u_j'\xt}{x^{II}_k}
     \nonumber \\
    & +u_i'\xo  \pdr{u_k'\xt \buj \xt}{x^{II}_k} - u_i'\xo  \pdr{R_{jk}}{x^{II}_k} = -\frac{1}{\rho}u_i'\xo \pdr{p'\xt}{x^{II}_k}\delta_{jk}
     \nonumber \\ 
    & + \nu u_i'\xo  \pdr{}{x^{II}_k}\pdr{{u_j'\xt}}{x^{II}_k} + g_j \beta u_i'\xo {T'\xt} ~.
     \label{urbc14}
 \end{align}

Eqn \rf{urbc13} and \rf{urbc14} represents the turbulence stress transport at two different points \textbf{$x^I$} and \textbf{$x^{II}$}. Combining Eqn \eqref{urbc13} and \eqref{urbc14} and averaging them gives the transport equation for two-point correlation ${Q_{ij}}$. The corresponding terms are added and transformed into scale space as discussed below.

Combining the first terms from Eqn \rf{urbc13} and \rf{urbc14} and average them as,
 \begin{align}
     & \ol{u_j'\xt\pdr{u_i'\xo}{t} +  u_i'\xo \pdr{u_j'\xt}{t}} \nonumber \\
     &= \pdr{}{t}\ol{\lrs{u_i'\xo u_j'\xt}}=\pdr{Q_{ij}(x^I, x^{II} )}{t}.
    \label{burbc15}
 \end{align}
 Applying the coordinate transformation on Eqn \rf{burbc15} using \rf{trans} - \rf{transf},
 \begin{equation}
   \pdr{Q_{ij}(\bm{x}, \bm{x+r})}{t}= \pdr{}{t}\ol{\lrs{u_i'(\bm{x}) u_j'(\bm{x+r})}}.
    \label{urbc15}
 \end{equation}
 Eqn \rf{urbc15} denotes the unsteady term.  The trace of the tensor divided by 2, $Q_{ii}(\bm{x}, \bm{x+r})/2$, represents the scale kinetic energy evolution with time. In the one-point limit, $|\bm{r}| \to 0$, it becomes the unsteady term (time evolution) of the average one-point turbulent kinetic energy. On the other limit, as $|\bm{r}| \to \infty$, $Q_{ii} =0$ i.e., the two points are uncorrelated. For a reasonable separation distance $|\bm{r}|$, $Q_{ii}(\bm{x}, \bm{x+}r) $ represents the part of turbulent kinetic energy with scales greater than $|\bm{r}|$. Hence, Eqn \rf{urbc15} represents the unsteady part of the evolution of scale kinetic energy.
 \\
 
Combining the second terms from Eqn \rf{urbc13} and \rf{urbc14} and average them as,
\begin{align}
& \ol{u_j'\xt \pdr{}{x^I_k}\lr{\buk \xo u_i'\xo}} + \ol{u_i'\xo  \pdr{}{x_k ^{II}}\lr{\buk \xt u_j'\xt}}
 \nonumber  \\
   = &\buk \xo \ol{\pdr{}{x^I_k}\lr{u_i'\xo u_j'\xt} }-  \buk \xo \ol{u_i'\xo \pdr{u_j'\xt}{x^I_k}} 
   \nonumber \\
   &+ \buk \xt \ol{\pdr{}{x_k ^{II}}\lr{u_i'\xo u_j'\xt} } - \buk \xt \ol{u_j'\xt\pdr{u_i'\xo}{x_k ^{II}}} 
   \nonumber \\ 
  = &  \buk \xo \pdr{Q_{ij}}{x^I_k} + \buk\xt \pdr{Q_{ij}}{x_k ^{II}}.
  \label{burbc16}
\end{align}
Note Eqn \rf{cross} is used here for simplification. Eqn \rf{burbc16} represents the convection of scale turbulence at two points. Applying the coordinate transformation on Eqn \rf{burbc16} using \rf{trans} - \rf{transf},
\begin{align}
 &  \buk \xo \pdr{Q_{ij}}{x^I_k} + \buk\xt \pdr{Q_{ij}}{x_k ^{II}} 
 \nonumber \\
 & =  \buk \x \pdr{Q_{ij}}{x_k} - \buk \x \pdr{Q_{ij}}{r_k} + \buk\xtt \pdr{Q_{ij}}{r_k}
 \nonumber \\
 & =  \buk \x \pdr{Q_{ij}}{x_k} - \pdr{Q_{ij}}{r_k}\lrs{ \buk \x - \buk\xtt} .
  \label{urbc16}
\end{align}
The first term on the right-hand side (RHS) of Eqn \rf{urbc16} denotes the convection of turbulence by the mean flow. The second term of Eqn \rf{urbc16} denotes the transport of turbulence in the scale space due to the mean velocity variation. 
In one point limit ($|\bm{r}| \to 0$), the second term goes to zero and only the convective term remains. On the other limit ($|\bm{r}| \to \infty$), both terms will be zero as the correlation becomes zero.
\\

Combining the third terms from Eqn \rf{urbc13} and \rf{urbc14}, the triple correlation terms of the fluctuations, and average them as,
\begin{align}
& \ol{u_j'\xt \pdr{}{x^I_k}\lr{u'_k\xo u_i'\xo}} + \ol{u_i'\xo \pdr{}{x_k ^{II}}\lr{u'_k\xt u_j'\xt}}
 \nonumber  \\
 = &  \ol{ \pdr{}{x^I_k}\lr{u_i'\xo u_k' \xo u_j'\xt} } + \ol{ \pdr{}{x_k ^{II}}\lr{u_k' \xt u_i'\xo u_j'\xt} } 
  \nonumber  \\ 
  &- \ol{u_i'\xo u_k' \xo \pdr{}{x^I_k}\lr{u_j'\xt} } - \ol{u_j'\xt u_k' \xt
  \pdr{}{x_k ^{II}}\lr{ u_i'\xo } } 
   \nonumber \\ 
  = &  \ol{ \pdr{}{x^I_k}\lr{u_i'\xo u_k' \xo u_j'\xt} } + \ol{ \pdr{}{x_k ^{II}}\lr{u_k' \xt u_i'\xo u_j'\xt} }  .
  \label{burbc17}
\end{align}
Note Eqn \rf{cross} is used here for simplification. Eqn \rf{burbc17} represents the triple correlation terms between quantities points at \textbf{$x^{I}$} and \textbf{$x^{II}$} Applying the coordinate transformation on Eqn \rf{burbc17} using \rf{trans} - \rf{transf},
\begin{align}
 &  \ol{ \pdr{}{x_k}\lr{u_i'\x u_k' \x u_j'\xtt} } - \ol{ \pdr{}{r_k}\lr{u_i'\x u_k' \x u_j'\xtt} } 
 \nonumber \\
 & + \ol{ \pdr{}{r_k }\lr{u_k' \xtt u_i'\x u_j'\xtt} } 
  \nonumber \\
  & = \pdr{}{x_k} \underbrace{\lr{\ol{u_i'\x u_k' \x u_j'\xtt}}}_{T_{ijk}} \nonumber \\
  & \textcolor{black}{- \ol{ \pdr{}{r_k}\lr{u_i'\x u_j'\xtt \lrs{ u_k' \x -  u_k' \xtt} } }}  \nonumber \\
  & = \pdr{T_{ijk}}{x_k}  - \ol{ \pdr{}{r_k}\lr{u_i'\x u_j'\xtt \lrs{ u_k' \x -  u_k' \xtt} } } .
  \label{urbc17}
\end{align}
Both terms in Eqn \rf{urbc17} are third-order moments of the fluctuating velocities. Third-order moments, in general, represent a transport due to the interaction among different scales. The first term, the gradient of $ T_{ijk} $, denotes a triple correlation term, which can be interpreted as the transport of turbulence in physical space because of the interaction between scales or the transport of scale turbulence $Q_{ij}$ due to the variation fluctuating velocity component $u'_k(x)$ in physical space. The second term, on the other hand, represented the transport of turbulence in the scale space due to the fluctuating velocity component. In the one-point limit ($|\bm{r}| \to 0 $), the second term goes to zero, and only the triple correlation term remains.
\\

Combining the fourth terms from Eqn \rf{urbc13} and \rf{urbc14} and average them as,
\begin{align}
    &\ol{u_j'\xt u_k'\xo \pdr{\bui\xo}{x^I_k} + u_i'\xo u_k'\xt \pdr{\buj \xt}{x_k ^{II}}}
    \nonumber \\
= & Q_{kj}\pdr{\bui\xo}{x^I_k} + Q_{ik}\pdr{\buj \xt}{x_k ^{II}} ~.
     \label{burbc18}
 \end{align}
 Eqn \rf{burbc18} represents the production of turbulence from the mean flow field at two different points. Applying the coordinate transformation on Eqn \rf{burbc18} using \rf{trans} - \rf{transf},
\begin{align}
  & Q_{kj}\pdr{\bui\xo}{x^I_k} + Q_{ik}\pdr{\buj \xt}{x_k ^{II}} 
  \nonumber \\
  &= Q_{kj}\pdr{\bui\x}{x_k} - Q_{kj}\pdr{\bui\x}{r_k} + Q_{ik}\pdr{\buj \xtt}{r_k}
  \nonumber \\
  & = Q_{kj}\pdr{\bui\x}{x_k} + Q_{ik}\pdr{\buj \xtt}{r_k} ~.
     \label{urbc18}
 \end{align}
The first term on Eqn \rf{urbc18} denotes the production of turbulence due to the mean flow gradient in physical space, whereas the second term denotes the production of turbulence due to the mean flow gradient in the scale space. \textcolor{black}{Note that the term $Q_{kj}\pdr{\bui\x}{r_k}$ vanishes as properties at $\mbx$ do not depend on $\mbr$}. In one point limit ($|\bm{r}| \to 0$), the second term becomes $Q_{ik}\pdr{\buj \x}{r_k}=0$ since the derivatives of quantities at $\bm{x}$ do not depend on $\bm{r}$. The fifth term from Eqn \rf{urbc13} and \rf{urbc14} vanishes upon averaging.

Combining the first term on the RHS of Eqn \rf{urbc13} and \rf{urbc14} and averaging them as,
\begin{align}
   & - \frac{1}{\rho} \ol{ u_j'\xt \pdr{p'\xo}{x^I_k} } \delta_{ik} 
  -  \frac{1}{\rho} \ol{ u_i'\xo \pdr{p'\xt}{x^{II}_k}} \delta_{jk}
  \nonumber \\
  & = - \frac{1}{\rho}\lrs{ \ol{ u_j'\xt \pdr{p'\xo}{x^I_i} } 
  + \ol{ u_i'\xo \pdr{p'\xt}{x^{II}_j}}} ~.
     \label{burbc19}
\end{align} 
 Eqn \rf{burbc19} represents the transport due to pressure at two different points. Applying the coordinate transformation on Eqn \rf{burbc19} using \rf{trans} - \rf{transf},
 \begin{align}
    - \frac{1}{\rho} & \ol{ u_j'\xtt \pdr{p'\x}{x_i}} - \frac{1}{\rho}\ol{ u_j'\xtt \pdr{p'\x}{r_i} } 
   \nonumber \\
   & +  \frac{1}{\rho}\ol{ u_i'\x \pdr{p'\xtt}{r_j}}
   \nonumber \\
   & =  - \frac{1}{\rho}\lrs{ \ol{ u_j'\xtt \pdr{p'\x}{x_i}} + \ol{ u_i'\x \pdr{p'\xtt}{r_j}}} .
     \label{urbc19}
\end{align} 
The first and second terms on Eqn \rf{urbc19} denote the transport due to pressure in physical space and scale space, respectively.

Combining the second term on the RHS of Eqn \rf{urbc13} and \rf{urbc14} containing viscosity and average them as,
\begin{align}
\nu~terms=  & \nu \ol{u_j'\xt \pdr{}{x^I_k}\pdr{u_i'\xo}{x^I_k}}
   +\nu \ol{u_i'\xo \pdr{}{x^{II}_k}\pdr{u_j'\xt}{x^{II}_k}} .
\end{align} 
These two terms are the viscous diffusion transport terms at two points. Applying the coordinate transformation using \rf{trans} - \rf{transf},
\begin{align}
 & \nu \ol{u_j'\xtt \pdr{}{x_k}\pdr{u_i'\xo}{x_k}}
   +\nu \ol{u_i'\x \pdr{}{r_k}\pdr{u_j'\xt}{r_k}} 
   \nonumber \\
   = &  \nu \ol{ \pdr{}{x_k} \lr{ u_j'\xtt \pdr{u_i'\x}{x_k}}}  - \nu \ol{ \pdr{u_i'\x}{x_k} \pdr{u_j'\xtt}{x_k}} 
   \nonumber \\ 
   &+ \nu \ol{ \pdr{}{r_k} \pdr{}{r_k}\lr{u_i'\x u_j'\xtt }}
   ~.
     \label{urbc20}
\end{align} 
 The first term on Eqn \rf{urbc20} represents the viscous diffusion transport in physical space, the second term is the dissipation rate, and the third term denotes the diffusion in the scale space. \textcolor{black}{In a single-point limit, as $|\bm{r}| \to 0$, the third term goes off, and only the first two terms ( viscous diffusion in physical space and dissipation) will remain.}
\\
 
Combining the last term on the RHS of Eqn \rf{urbc13} and \rf{urbc14} and average them as,
\begin{align}
   & g_i \beta \ol{u_j'\xt{T' \xo}} + g_j \beta \ol{u_i'\xo{T' \xt}}
   \nonumber \\
   & =\beta \lr{g_i\ol{u_j'\xt{T' \xo}} + g_j \ol{u_i'\xo{T' \xt}}} ~.
     \label{burbc21}
\end{align} 
 Applying the coordinate transformation on Eqn \rf{burbc21} using \rf{trans} - \rf{transf},
\begin{align}
  & \beta \lr{ g_i\ol{u_j'\xtt{T' \x}} + g_j \ol{u_i'\x{T' \xtt}}}
  \nonumber \\
  &= \beta \lr{g_i H_j' + g_j H_i}  ~,
     \label{urbc21}
\end{align} 
where $H_i=\ol{u_i'\x{T' \xtt}}$, and $H'_j=\ol{u_j'\xtt{T' \x}}$ are the turbulent scale heat flux. Eqn \rf{urbc21} denotes the production of $Q_{ij}$ due to the source(buoyancy). \textcolor{black}{In one-point limit, it will become $ \beta (g_i \av{u_j'(x)T'(x)} +g_j\av{u_i'(x)T'(x)}) $, heat-flux production due to buoyancy.} 
\\

Combining Eqn \rf{urbc15}, \rf{urbc16}, \rf{urbc17},  \rf{urbc18},  \rf{urbc19}, \rf{urbc20}, and \rf{urbc21} and rearranging,
\begin{align}
& \pdr{Q_{ij}(\bm{x},\bm{r})}{t} + \buk \x \pdr{Q_{ij}(\bm{x},\bm{r})}{x_k} = - Q_{kj}(\bm{x},\bm{r})\pdr{\bui \x }{x_k} 
    \nonumber \\
    & - \pdr{T_{ijk}(\bm{x},\bm{r})}{x_k} +  \beta \lr{g_j H_i + g_i H_j'}
    - \frac{1}{\rho}\ol{ u_j'\xtt \pdr{p'\x}{x_k}{\delta_{ik}} } 
     \nonumber \\
    &+ \nu \pdr{}{x_k}\lr{\ol{u_j'\xtt \pdr{u_i'\x}{x_k}}} - \nu \ol{\pdr{u_j'\xtt }{x_k}\pdr{u_i'\x}{x_k} } 
    \nonumber \\
    & - Q_{ik}(x,r)\pdr{\buj \xtt}{r_k} - \frac{1}{\rho}\ol{ u_i'\x\pdr{p'\xtt}{r_k}{  \delta{jk}}} 
    \nonumber \\
    & + \nu \pdr{}{r_k}\pdr{Q_{ij}}{r_k} 
    + \pdr{}{r_k}\lr{Q_{ij}\lrs{\buk \x - \buk \xtt} } 
    \nonumber \\
    & + \pdr{}{r_k}\lr{\av{u_i'\x u_j'\xtt \lrs{u_k'\x - u_k'\xtt}}} ~, 
    \label{uu44}
\end{align} 
we get a general two-point velocity correlation transport equation for incompressible, inhomogeneous, and anisotropic turbulent flow with a buoyancy source term. 

The scale kinetic energy transport equation can be extracted from \rf{uu44} by simply taking the traces of turbulent stress tensor, i.e. by putting $i=j$,
\begin{align}
    & \pdr{Q_{ii}(\bm{x},\bm{r})}{t} + \buk \x \pdr{Q_{ii}(\bm{x},\bm{r})}{x_k} = \underbrace{- Q_{ki}(\bm{x},\bm{r})\pdr{\bui \x }{x_k}}_{\bm{P}_{x}}
    \nonumber \\
    & - \underbrace{\pdr{T_{iik}(\bm{x},\bm{r})}{x_k}}_{\bm{T}} + \underbrace{g_i \beta \lr{H_i + H_i'}}_{\bm{S}}
    - \underbrace{\frac{1}{\rho}\ol{ u_i'\xtt \pdr{p'\x}{x_k}{\delta_{ik}}} }_{\bm{D}_x^p} 
     \nonumber \\
    & + \underbrace{\nu \pdr{}{x_k}\lr{\ol{u_i'\xtt \pdr{u_i'\x}{x_k}}}}_{\bm{D}_x^v} - \textcolor{black}{\underbrace{\nu \ol{\pdr{u_i'\xtt }{x_k}\pdr{u_i'\x}{x_k} }}_{\bm{\epsilon}} }
    \nonumber \\
    & - \underbrace{Q_{ik}(x,r)\pdr{\bui \xtt}{r_k}}_{\bm{P}_r} -\underbrace{\frac{1}{\rho}\ol{ u_i'\x\pdr{p'\xtt}{r_k}{  \delta{ik}}}}_{\bm{D}_r^p} 
    \nonumber \\
    & + \underbrace{\nu \pdr{}{r_k}\pdr{Q_{ii}}{r_k}}_{\bm{\mathcal{D}}_r} 
    + \underbrace{\pdr{}{r_k}\lr{Q_{ii}\lrs{\buk \x - \buk \xtt} }}_{\bm{SST}_{m}} 
    \nonumber \\
    & + \underbrace{\pdr{}{r_k}\lr{\av{u_i'\x u_i'\xtt \lrs{u_k'\x - u_k'\xtt}}}}_{\bm{SST}_{f}} ~.
     \label{uu44_ii}
\end{align}

The terms on the LHS of Eqn \rf{uu44_ii} are the unsteady and convective terms of turbulence kinetic energy. On the RHS: $\bm{P}_x$ is the production of turbulence due to the mean velocity gradient in physical space, $\bm{T}$ is the triple correlation term which is the transport of kinetic energy due to the fluctuating velocity field, $\bm{S} $ is the source term or production of turbulence due to buoyancy, $\bm{D}_x^p $ represents the transport due to pressure in physical space, $\bm{D}_x^v $ represents the transport in physical space due to viscosity, $\bm{\epsilon} $ represents the dissipation rate, $\bm{P}_r $ represents the production due to the mean velocity gradient in scale space, $\bm{D}_r^p $ is the transport due to pressure in scale space, $\bm{\mathcal{D}}_r $ is the diffusion in scale space due to viscosity, $\bm{SST}_{m} ~\text{and}~ \bm{SST}_{f} $ are the transport terms in scale space or interscale transport term due to the mean and fluctuating velocity fields, respectively.

 Eqn \rf{uu44_ii} is the general two-point transport equation of scale kinetic energy for incompressible, anisotropic, and inhomogeneous turbulence. In one point limit, Eqn \rf{uu44_ii} reduces into the conventional (one-point) kinetic energy transport equation, which is written as,
\begin{align}
    & \pdr{\av{u'_iu'_i}}{t} + \av{u_k}\pdr{\av{u'_iu'_i}}{x_k} = -\av{u'_iu'_k}\pdr{\av{u_i}}{x_k} - \pdr{\av{u'_iu'_iu'_k}}{x_k} + 2g_i\beta \av{u'_iT'} 
    \nonumber \\ 
    & -\frac{1}{\rho} \pdr{\av{u'_ip'}}{x_i} + \nu \av{\pdr{}{x_k}\lr{u'_i\pdr{u'_i}{x_k} }} - \nu \av{\pdr{u'_i}{x_k}\pdr{u'_i}{x_k}}. 
    \label{ke-1}
\end{align}
Eqn \rf{ke-1} governs the transport of turbulence kinetic energy in physical space. The terms on the right-hand side result in energy production, transport due to the fluctuating field, production of turbulence due to buoyancy, transport due to pressure, transport due to viscosity, and the dissipation rate. Hence, the remaining terms in Eqn \rf{uu44_ii} ($\bm{P}_r,~ \bm{D}^p_r,~ \bm{D}_r, ~ \bm{SST}_{m},~ \text{and} \bm{SST}_{f} $) corresponds to the transport of kinetic energy in the scale space. Note again that the scale kinetic energy we defined as $Q_{ii}=\ol{u'_i\xo u'_i\xt }$ is not the energy density or the energy at a specific scale $|\bm{r}|$. The notation $Q_{ii} $ denotes the part of turbulence energy with scales larger than $|\bm{r}|$. In section \rf{density}, we will define the energy density, which is the equivalent of energy density in wavenumber space for homogeneous turbulence, and derive its transport equation to study the energy budget at each scale.
\section{Scale heat flux transport equation} \label{sec:heat}
 Similar to the two-point velocity-velocity correlation as discussed above, heat transport can be measured using a velocity-temperature correlation. 
 Two-point velocity-temperature correlation for incompressible flow is defined as,
\begin{align}
H_{i }(x^I, x^{II})= \ol{u_i'\xo T'\xt} ~.
\label{rbc10} 
\end{align}
When $|\bm{r}|=0$, $H_i(x)$ becomes the local turbulent heat flux along $i$-direction at a point $\bm{x}$ in physical space. Based on the same argument we presented in section \rf{sec-Qii} for $Q_{ii}$, $H_i(\bm{x},\bm{r})$ represents the part of turbulent heat flux for a scale greater than $|\bm{r}|$. 
The evolution of the unaveraged velocity-temperature correlation, $h_i=u_i'\xo T'\xt$, in time and space can be written as,


\begin{align}
   \pdr{h_i}{t} + \buk  \frac{\partial h_i}{\partial x_k} 
   & = T'\xt \lr{\pdr{u_i'\xo}{t} + \buk \xo\pdr{u_i'\xo}{x^I_k}} 
   \nonumber \\
   &+  u_i'\xo \lr{ \pdr{T' \xt}{t} + \buk \xt \pdr{T'\xt}{x_k ^{II}}} ~.
    \label{rbc12}
\end{align}

Comparing Eqn \rf{rbc12} with Eqn \rf{uu12}, and Eqn \rf{t20}, a two-point heat flux transport equation can be derived by multiplying Eqn \rf{uu12} by $ T'\xt$ , and Eqn \rf{t20} by ${u_i'\xo}$ as,


\begin{align}
    & T'\xt\pdr{u_i'\xo}{t} + T'\xt \pdr{\buk \xo u_i'\xo}{x^I_k} +  T'\xt \pdr{u'_k\xo u_i'\xo}{x^I_k}
    \nonumber \\
    &   + T'\xt  \pdr{u_k'\xo \bui \xo}{x^I_k}  - T'\xt \pdr{R_{ki}}{x^I_k} = -\frac{1}{\rho}T'\xt\pdr{p'\xo}{x^I_k}\delta_{ik} 
    \nonumber \\
    & + \nu T'\xt \pdr{}{x^I_k}\pdr{{u_i'\xo}}{x^I_k} + g_i \beta T'\xt{T'\xo} ~,~\text{and}
     \label{rbc13}
 \end{align}
\begin{align}
    & u_i'\xo \pdr{T'\xt}{t} +  u_i'\xo \pdr{\buk \xt T'\xt}{x_k ^{II}} + u_i'\xo \pdr{u'_k\xt T'\xt}{x_k ^{II}} 
    \nonumber \\
    & + u_i'\xo u_k'\xt \pdr{\bT\xt}{x_k ^{II}}  - u'_i\xo \ol{\pdr{E_k}{x_k}}  = \alpha u_i'\xo\pdr{}{x_k ^{II}}\pdr{T'\xt}{x_k ^{II}}. 
    \nonumber \\ 
    &
    \label{rbc14}
 \end{align}
 Combining the first terms of Eqn \rf{rbc13} and \rf{rbc14} and averaging it as,
 \begin{align}
     \ol{T'\xt\pdr{u_i'\xo}{t} +  u_i'\xo \pdr{T'\xt}{t}}= \pdr{}{t}\ol{\lrs{u_i'\xo T'\xt}},
    \label{brbc15}
 \end{align}
followed by the coordinate transformation on Eqn \rf{brbc15} using \rf{trans} - \rf{transf} results in,
\begin{align}
     & \pdr{}{t}\ol{\lrs{u_i'\x T'\xtt}} =\pdr{H_i(\bm{x},\bm{r})}{t}.
    \label{rbc15}
 \end{align}
Eqn \rf{rbc15} is the unsteady part in the transport equation, which represents the time evolution of the average scale heat flux.\\
 
  Combining the second terms of Eqn \rf{rbc13} and \rf{rbc14}, and averaging,
\begin{align}
 &  \ol{ T' \xt \pdr{}{x^I_k} \lr{\ol{u_k} \xo u_i'\xo}
  }  + \ol{u_i'\xo  \pdr{}{x_k^{II}} \lr{\buk \xt T'\xt} } 
  \nonumber \\  
   = &\buk \xo \ol{\pdr{}{x^I_k}\lr{u_i'\xo T'\xt} }-  \buk \xo \ol{u_i'\xo \pdr{T'\xt}{x^I_k}} 
   \nonumber \\
   & + \buk \xt \ol{\pdr{}{x_k ^{II}}\lr{u_i'\xo T'\xt} } - \buk \xt \ol{T'\xt\pdr{u_i'\xo}{x_k ^{II}}} 
  \nonumber  \\
  = & \buk \xo \pdr{H_i(x^I, x^{II}) }{x^I_k} + \buk\xt \pdr{H_i(x^I, x^{II}) }{x_k ^{II}} ~.
       \label{brbc16}
\end{align}
The cross derivatives given in Eqn \rf{cross} are used in the above reduction. Eqn \rf{brbc16} represents the convection of heat flux by the mean flow field at two points. Applying coordinate transformation on Eqn \rf{brbc16} using \rf{trans} - \rf{transf},
\begin{align}
   & \buk \x \pdr{H_i(\bm{x},\bm{r}) }{x_k} - \buk \x \pdr{H_i(\bm{x},\bm{r}) }{r_k}  + \buk\xtt \pdr{H_i(\bm{x},\bm{r})}{r_k} 
   \nonumber \\
   & = \buk \x \pdr{H_i(\bm{x},\bm{r}) }{x_k} - \pdr{H_i(\bm{x},\bm{r}) }{r_k}\lrs{\buk \x - \buk\xtt }   ~.
       \label{rbc16}
\end{align}
The first term on Eqn \rf{rbc16} denotes the convection of scale heat flux in physical space due to the mean velocity field, and the second term denotes the transport of scale heat flux in the scale space due to the mean velocity differences. In the one-point limit ($|\bm{r}| \to 0$), the second term goes to zero, and only the convection of turbulent heat flux in physical space remains.
\\

 Combining the third terms of Eqn \rf{rbc13} and \rf{rbc14}, utilising Eqn \rf{cross} and averaging,
\begin{align}
 &  \ol{ T'\xt \pdr{}{x^I _k}\lr{u'_k\xo u_i'\xo} } + \ol{ u_i'\xo \pdr{}{x_k^{II}}\lr{u'_k\xt T'\xt} } 
  \nonumber \\  
   = & \ol{ \pdr{}{x^I_k}\lr{u_i'\xo u_k' \xo T'\xt} } - \ol{u_i'\xo u_k' \xo \pdr{T'\xt}{x^I_k}}
  \nonumber  \\ 
   & + \ol{ \pdr{}{x_k ^{II}}\lr{u_k' \xt u_i'\xo T'\xt} } - \ol{T'\xt u_k' \xt \pdr{ u_i'\xo}{x_k ^{II}} } 
   \nonumber \\  
   = & \ol{ \pdr{}{x^I_k}\lr{u_i'\xo u_k' \xo T'\xt} }  + \ol{ \pdr{}{x_k ^{II}}\lr{u_k' \xt u_i'\xo T'\xt} } 
   ~.
       \label{brbc160}
\end{align}
Eqn \rf{brbc160} represents the triple correlation term between two points. Applying the coordinate transformation on Eqn \rf{brbc160} using \rf{trans} - \rf{transf}, 
\begin{align}
 & \ol{ \pdr{}{x_k}\lr{u_i'\x u_k' \x T'\xtt} } -  \ol{ \pdr{}{r_k}\lr{u_i'\x u_k' \x T'\xtt}} 
 \nonumber \\
 & + \ol{ \pdr{}{r_k} \lr{u_k' \xtt u_i'\x T'\xtt} } =
 \ol{ \pdr{}{x_k}\lr{u_i'\x u_k' \x T'\xtt} }
 \nonumber \\
 &  -  \ol{ \pdr{}{r_k}\lr{u_i'\x T'\xtt \lrs{ u_k' \x-  u_k' \xtt}}} 
 \nonumber \\
 &= \pdr{\mathcal{T}_{ik}}{x_k} -  \ol{ \pdr{}{r_k}\lr{u_i'\x T'\xtt \lrs{ u_k' \x-  u_k' \xtt}}} ~.       \label{rbc160}
\end{align}
The first term on Eqn \rf{rbc160} denotes the triple correlation term, which is a third-order moment of the fluctuating quantities representing the transport of scale heat flux due to the interaction among scales. Another way of interpreting the triple correlation term is that it represents the transport of scale heat flux $u_i' \x T'\xtt$ along $i$-direction due to the $k^{th}$ component of fluctuating velocity, $u_k'\x$. The second term on Eqn \rf{rbc160} denotes the transport of heat flux in the scale space due to the variations in fluctuating velocity. In the one-point limit ($|\bm{r} |\to 0$), the second term goes to zero, and only the triple-correlation term remains.
\\

Collecting the fourth terms of Eqn \rf{rbc13} and \rf{rbc14} and averaging,
 \begin{align}
    &\ol{T'\xt u_k'\xo \pdr{\bui\xo}{x^I_k} + u_i'\xo u_k'\xt \pdr{\bT\xt}{x_k ^{II}}}
    \nonumber \\
= & \ol{T'\xt u_k'\xo }\pdr{\bui\xo}{x^I_k} + \ol{u_i'\xo u_k'\xt}\pdr{\ol{T}\xt}{x_k ^{II}} ~. 
\label{brbc17}
 \end{align}
The first term on Eqn \rf{brbc17} denotes the production of turbulent heat flux due to the mean velocity at the first point, while the second term is the production of turbulent stress due to the mean temperature at the second point. Applying the coordinate transformation on Eqn \rf{brbc17} using \rf{trans} - \rf{transf}, 
 \begin{align}
 & \ol{T'\xtt u_k'\x }\pdr{\bui \x}{x_k} -  \ol{T'\xtt u_k'\x }\pdr{\bui \x}{r_k}
 \nonumber \\
 & + \ol{u_i'\x u_k'\xtt}\pdr{\ol{T}\xtt}{r_k} 
 \nonumber \\
 & = \ol{T'\xtt u_k'\x }\pdr{\bui \x}{x_k}  + \ol{u_i'\x u_k'\xtt}\pdr{\ol{T}\xtt}{r_k} 
  \nonumber \\
 & = H_k(\bm{x},\bm{r}) \pdr{\bui \x}{x_k}  + Q_{ik}(\bm{x},\bm{r})\pdr{\ol{T}\xtt}{r_k} 
 ~.  \label{rbc17} 
 \end{align}
The first term on Eqn \rf{rbc17} represents the production due to the mean velocity gradient in physical space, whereas the second term denotes the production due to the mean temperature gradient in scale space. 
 \\
  
  Collecting RHS of equation \rf{rbc13} and \rf{rbc14} and averaging,
\begin{align}
  & RHS\rf{rbc13} + RHS\rf{rbc14}= 
  \nonumber \\
  &-\frac{1}{\rho} \ol{ T'\xt \pdr{p'\xo}{x^I_k}}\delta_{ik} + \nu \ol{T'\xt \pdr{}{x^I_k}\pdr{u_i'\xo}{x^I_k}} 
   \nonumber \\ 
   & + g_i \beta \ol{T'\xt{T'\xo}} + \alpha \ol{u_i'\xo\pdr{}{x_k ^{II}}\pdr{T'\xt}{x_k ^{II}}}
   ~.
     \label{brbc18}
\end{align}
The first term on Eqn \rf{brbc18} is the transport due to pressure, the second term is the transport due to viscous diffusion, the third term is the source(buoyancy) term, and the last term is the transport due to thermal diffusion. Applying the coordinate transformation on Eqn \rf{brbc18} using \rf{trans} - \rf{transf}, 
\begin{align}
  &-\frac{1}{\rho} \ol{ T'\xtt \pdr{p'\x}{x_k}}\delta_{ik} + \frac{1}{\rho} \ol{ T'\xtt \pdr{p'\x}{r_k}}\delta_{ik} 
  \nonumber \\
  &+ \nu \ol{T'\xtt \pdr{}{x_k}\pdr{u_i'\x}{x_k}} - \nu \ol{T'\xtt \pdr{}{r_k}\pdr{u_i'\x}{r_k}} 
   \nonumber \\ 
   & + g_i \beta \ol{T'\xtt{T'\x}} + \alpha \ol{u_i'\x\pdr{}{r_k }\pdr{T'\xtt}{r_k }}
   \nonumber \\
   =& -\frac{1}{\rho} \ol{ T'\xtt \pdr{p'\x}{x_k}}\delta_{ik}  + \nu \ol{T'\xtt \pdr{}{x_k}\pdr{u_i'\x}{x_k}}
   \nonumber \\
   &  + g_i \beta \ol{T'\xtt{T'\x}} + \alpha \ol{u_i'\x\pdr{}{r_k }\pdr{T'\xtt}{r_k }}
   \nonumber \\
   =& -\frac{1}{\rho} \ol{ T'\xtt \pdr{p'\x}{x_k}}\delta_{ik}  + \nu \ol{ \pdr{}{x_k}\lr{T'\xtt \pdr{u_i'\x}{x_k}}}
   \nonumber \\
   & - \nu \ol{ \pdr{T'\xtt }{x_k}\pdr{u_i'\x}{x_k}} +  \alpha \pdr{}{r_k }\pdr{}{r_k }\ol{\lr{u_i'\x T'\xtt} }
   \nonumber \\
   & + g_i \beta \ol{T'\xtt{T'\x}}
   ~.
     \label{rbc18}
\end{align}
The first term on Eqn \rf{rbc18} denotes the transport of turbulence heat flux due to pressure, the second term represents the diffusion in physical space due to viscosity, the third term is the dissipation rate, the fourth term denotes the diffusion in the scale space, and the last term is the source term, production of turbulence heat flux due to buoyancy. 

Combining Eqn \rf{rbc15}, Eqn \rf{rbc16},  Eqn \rf{rbc160}, Eqn \rf{rbc17}, and Eqn \rf{rbc18}, and rearranging terms we, get the two-point transport equation for heat flux as,
\begin{align}
    &  \pdr{H_i}{t} +  {\buk \x \pdr{H_i}{x_{k}}} =  \underbrace{g_i\beta \ol{T'\xtt{T'\x}}}_{\bm{S_\theta}} - \underbrace{H_k\pdr{\bui\x}{x_{k}}}_{\bm{P_{\theta x}}} 
    \nonumber \\
    & -\underbrace{Q_{ik}\pdr{\ol{T}\xtt}{r_{k}}}_{\bm{P_{\theta r}}} -  \underbrace{\pdr{\mathcal{T}_{ik}}{x_k}}_{\bm{T_\theta}} -  \underbrace{\frac{1}{\rho}{\ol{ T'\xtt\pdr{ p'\x}{x_{k}}}\delta_{ik}}}_{\bm{D^p_\theta}}
    \nonumber  \\ 
     &  +  \underbrace{\nu \ol{ \pdr{}{x_k}\lr{T'\xtt \pdr{{u_i'\x}}{x_{k}}}}}_{\bm{D^v_\theta}}
     -   \underbrace{\nu \ol{ \pdr{T'\xtt}{x_k}{ \pdr{{u_i'\x}}{x_{k}}}}}_{\bm{\epsilon_\theta}}
    \nonumber  \\ 
   & + \underbrace{\pdr{}{r_{k}}\lr{ \ol{u_i'\x  T'\xtt \lrs{u_k' \x -u_k' \xtt} }}}_{\bm{SST}_{\theta f}} 
   \nonumber \\
   & + \underbrace{\pdr{}{r_{k}}\lr{\lrs{\buk\x- \buk \xtt  }H_i}}_{\bm{SST}_{\theta m}} +  \underbrace{\alpha { \pdr{}{r_k}\lr{\pdr{H_i}{r_{k}}}} }_{\bm{\mathcal{D}}_{\theta r}} .
    \label{ut-rbc}
\end{align}

The two terms on the left-hand side of Eqn \rf{ut-rbc} are the unsteady and convective terms of scale heat-flux. On the right-hand side: $\bm{S_\theta} ~ \text{and} ~ \bm{P}_{\theta x} $ denote the production of scale heat flux due to buoyancy and mean flow field in physical space, respectively,  $\bm{P_{\theta r}} $ is the production due to mean temperature in scale space, $\bm{T_\theta} $ represents triple correlation, $\bm{D^p_\theta} ~\text{and} ~ \bm{D^v_{\theta}} $ denote the transport of scale heat-flux due to pressure and viscosity in physical space, $\bm{\epsilon_\theta} $ represents dissipation rate, $\bm{SST_{\theta f}} ~ \text{and} ~ \bm{SST_{\theta m}} $ denote the scale-space transport or interscale transport (scale-space dynamics due to inhomogeneity in the fluctuating and mean fields), and $\bm{\mathcal{D}_{\theta r}} $ denotes the diffusion of heat in scale space. 

Eqn \rf{ut-rbc} is the general two-point transport equation of heat flux for incompressible, anisotropic, and inhomogeneous turbulence with buoyancy. In one point limit, Eqn \rf{ut-rbc} reduces into the conventional one-point turbulent heat flux transport equation as,
\begin{align}
    &\pdr{\av{u'_iT'}}{t} + \av{u_k}\pdr{\av{u'_iT'}}{x_k} = \av{g_i\beta T'^2} - \av{u'_kT'}\pdr{\av{u_i}}{x_k} - \pdr{\av{u'_iu'_kT'}}{x_k} 
    \nonumber \\
    &- \frac{1}{\rho}\av{T'\pdr{p'}{x_i}} + \nu \av{\pdr{}{x_k}\lr{T'\pdr{u'_i}{x_k} }} - \nu \av{\pdr{u'_i}{x_k}\pdr{T'}{x_k} } ,
\end{align}
which governs the evolution of turbulent heat flux in the physical space. The terms on the right-hand side represent the production due to buoyancy and mean flow field, transport due to fluctuating field, transport due to pressure and viscosity, and the dissipation rate, respectively.  The remaining terms from Eqn \rf{ut-rbc}, $\bm{P\theta}_r,~ \bm{SST}_{\theta f},~ \bm{SST}_{\theta m},~\text{and} ~\bm{\mathcal{D}}_{\theta r} $, capture the transport of heat flux in the scale space of turbulence. To analyze the heat flux budget and the heat transfer at each scale, in the next section, we defined heat flux density, which is analogous to the heat flux density in wavenumber space for homogeneous turbulence.

\section{Energy and heat flux density in scale space}
\label{density}
Understanding turbulence primarily relies on understanding the energy distribution among eddies and their interactions.
For canonical homogeneous turbulence, the energy/heat budget of each eddie/scale has been studied in the wave number space using the Fourier transformation tool. However, we can not apply Fourier transfer and study the energy spectrum for inhomogeneous turbulent flows. 

In our discussion in the previous sections, we derived two transport equations for turbulence kinetic energy and heat flux. However, note that the quantities defined there ($Q_{ii} ~\text{and}~ H_i$) are not the energy and heat flux density. Rather, they represent the part of turbulence kinetic energy and heat flux for a scale greater than $|\bm{r}|$.

To obtain the exact kinetic energy and heat flux at each scale/eddy size, we introduce energy density and heat flux density, which are nothing but the $\bm{r}$ derivative of Eqn \rf{uu44} and Eqn \rf{ut-rbc}. Energy density and heat flux density are analogous to the energy spectrum and heat flux spectrum in the wave number space, but they are not restricted to homogeneous or isotropic cases.

The energy density and heat flux density are defined as,
\begin{equation}
    E(x,r_\alpha) = -\frac{1}{2} \pdr{Q_{ii}}{r_\alpha},~\text{and}
    \label{ed}
\end{equation}
\begin{equation}
    J_i(x,r_\alpha) = -\pdr{H_{i}}{r_\alpha}, \label{hf}
\end{equation}
where $\alpha, i$ are indices taking the values 1,2 or 3 depending on the direction.
\textcolor{black}{The energy(heat) density means the amount of energy(heat) carried by a single eddy of scale $r$. The definition of energy density has to satisfy two conditions: it must be positive and the integral of it over all the scales should give the kinetic energy at that point. If you use the definition of energy density \cite{hamba_2015} as provided here, the total turbulent kinetic energy at $\mbx$, $KE = \int^{\infty}_0 E(\mbx,\mbr) dr = \frac{1}{2}(Q_{ii}(\mbx,0) - Q_{ii}(\mbx,\infty))$. A similar analysis is applied to the heat flux density also.}

To obtain the energy density transport equation, we apply $-\frac{1}{2} \pdr{}{r_{\alpha}}\lrs{Eqn~\rf{uu44_ii}}$ and upon simplifying we get,

\begin{align}
    & {\pdr{E(\bm{x},r_\alpha)}{t}} + {\buk \x \pdr{E(\bm{x,}r_\alpha)}{x_k}} =  \underbrace{\frac{1}{2}\pdr{}{r_\alpha}\lr{\pdr{T_{iik}(\bm{x},\bm{r})}{x_k}}}_{\bm{T_{r_\alpha}}} 
    \nonumber \\
    &+ \underbrace{  {\frac{1}{2}\pdr{}{r_\alpha}\lr{ Q_{ki}(\bm{x},\bm{r})}\pdr{\bui \x }{x_k}}  + \frac{1}{2}\pdr{}{r_\alpha}\lr{Q_{ik}(\bm{x},\bm{r})\pdr{\bui \xtt}{r_k}}}_{\bm{P_{r_\alpha}}}
    \nonumber \\
    &  - \underbrace{\frac{1}{2} \beta g_i\pdr{}{r_\alpha}{ \lr{H_i(\bm{x},\bm{r}) + H_i'(\bm{x},\bm{r})}}}_{\bm{S_{r_\alpha}}}  
    \nonumber \\
    & - \underbrace{\frac{\nu}{2}\pdr{}{r_\alpha}\lr{\pdr{}{x_k}\lr{\ol{u_i'\xtt \pdr{u_i'\x}{x_k}}}}}_{\bm{D^v_{r_\alpha}}}
    \nonumber \\
    & + \underbrace{\frac{1}{2\rho}\pdr{}{r_\alpha}\lr{\ol{ \pdr{u_i'\xtt p'\x}{x_i}} + \ol{ \pdr{u_i'\x p'\xtt}{r_i}} }}_{\bm{D^p_{r \alpha}}} 
     \nonumber \\
    & + \underbrace{\frac{\nu}{2}\pdr{}{r_\alpha}\lr{ \ol{\pdr{u_i'\xtt }{x_k}\pdr{u_i'\x}{x_k} }}}_{\bm{\epsilon_{r_\alpha}}} + \underbrace{\nu{ \pdr{}{r_k}\pdr{E(\bm{x}, r_\alpha)}{r_k}}}_{\mathcal{D}_{r\alpha}} 
    \nonumber \\
    &    - \underbrace{\frac{1}{2}\pdr{}{r_\alpha}\lr{\pdr{}{r_k}\lr{Q_{ii}(\bm{x},\bm{r})\lrs{\buk \x - \buk \xtt} }}}_{\bm{SST_{r \alpha,m}}} 
    \nonumber \\
    & - \underbrace{\frac{1}{2}\pdr{}{r_\alpha}\lr{\pdr{}{r_k}\lr{\av{u_i'\x u_i'\xtt \lrs{u_k'\x - u_k'\xtt}}}}}_{\bm{SST_{r \alpha, f}}} ~.
     \label{uu46}
\end{align}
\textcolor{black}{The intermediate step is shown in the appendix Eqn \rf{uu45}.} The two terms on the left-hand side are the unsteady term and convective terms of energy density. On the right-hand side, $\bm{T_{r \alpha}} $ is the triple correlation, which is the kinetic energy at scale $r_\alpha$ due to the interaction among different scales\textcolor{black}{, is the transport of energy density by turbulent field}. $\bm{P_{r \alpha}} $ is the production of turbulence energy at scale $r_\alpha$ due to mean flow gradient in physical space and scale space, $\bm{S_{r \alpha}} $ denotes buoyancy source term\textcolor{black}{, which is the energy density added or extracted due to buoyancy at scale $r_{\alpha}$}, $\bm{D^p_{r \alpha}} $ represents the pressure velocity correlation contribution to the scale energy budget, \textcolor{black}{ the energy transported by pressure at scale $r_{\alpha}$}, $\bm{D^v_{r \alpha}} $ represents the viscous diffusion term, $\bm{\epsilon_{r \alpha}} $ denotes the energy dissipation rate, \textcolor{black}{or the energy density dissipated due to viscous effects at scale $r_{\alpha}$}, $\bm{\mathcal{D}_{r \alpha}}$ represents the diffusion term in scale space, $\bm{SST_{r \alpha, m}}$ and $\bm{SST_{r \alpha, f}} $ denotes the interscale transport terms due to the mean and fluctuating flow field, respectively. 

Eqn \rf{uu46} is the transport equation for the kinetic energy density, the energy transport at each scale $\bm{r}$. It depends on the scale $|\bm{r}|$ and the direction $\alpha$. This indicates that eddies of the same size/scale $|\bm{r}|$ could have different energy budgets depending on their orientation or direction. Thus, the energy density obtained is for a general separation vector $\mathbf{r}$, and a one-dimensional energy density function can be obtained by integrating over a spherical shell of radius of $\lvert \mbr \rvert$ as,
\begin{equation}
    E(\mbx,\lvert\mbr\rvert) = \oint E(\mbx,\mbr) d\phi d\eta,
    \label{1d_rep}
\end{equation}
where $r_1 = \lvert \mbr \rvert sin \phi cos \eta$, $r_2 = \lvert \mbr \rvert sin \phi sin \eta$, and $r_3 = \lvert \mbr \rvert cos \phi$. The term $E(\lvert\mbr\rvert)d\mbr$ is the amount of energy contained in eddies of size $\lvert\mbr\rvert$ to $\lvert \mbr + d\mbr \rvert$, which is similar to the relation $E(k) dk$ in the Fourier representation in homogenous turbulence.

The heat flux density transport equation can be obtained in a similar way by taking the $r$ derivative of Eqn \rf{ut-rbc} and further simplification as, 
\begin{align}
    &  \pdr{J_i(x,r_\alpha)}{t} +  {\buk \x \pdr{J_i(x,r_\alpha)}{x_{k}}} = - \underbrace{\pdr{}{r_\alpha}\lr{g_i\beta \ol{T'\xtt{T'\x}}}}_{\bm{Sr_\alpha}} 
    \nonumber \\
    & \underbrace{ + \pdr{H_k}{r_\alpha}{\pdr{\bui\x}{x_{k}}}  + \pdr{}{r_\alpha}\lr{Q_{ik}\pdr{\ol{T}\xtt}{r_{k}}}}_{\bm{P\theta_{r\alpha}}} +  \underbrace{\pdr{}{r_\alpha}{\pdr{\mathcal{T}_{ik}}{x_k}}}_{T\theta_{r\alpha}}
    \nonumber \\
    &  +  \underbrace{\frac{1}{\rho}\pdr{}{r_\alpha}\lr{\ol{ T'\xtt\pdr{ p'\x}{x_{k}}}\delta_{ik}}}_{\bm{D\theta^p_{r\alpha}}}  \nonumber \\
    &-  \underbrace{\nu \pdr{}{r_\alpha}\lr{ \ol{ \pdr{}{x_k}\lr{T'\xtt \pdr{{u_i'\x}}{x_{k}}}}}}_{\bm{D\theta^v_{r\alpha}}} \nonumber \\
     &+   \underbrace{\pdr{}{r_\alpha}\lr{ \ol{ \pdr{T'\xtt}{x_k}{ \pdr{{u_i'\x}}{x_{k}}}}}}_{\bm{\epsilon\theta_{r\alpha}}} +  \underbrace{\alpha {\pdr{}{r_k}\lr{\pdr{J_i(x,r_\alpha)}{r_{k}}}} }_{\bm{\mathcal{D}\theta_{r\alpha}}}
    \nonumber  \\ 
   & - \underbrace{\pdr{}{r_\alpha}\pdr{}{r_{k}}\lr{ \ol{u_i'\x  T'\xtt \lrs{u_k' \x -u_k' \xtt} }}}_{\bm{SST\theta}_{r_\alpha f}} 
   \nonumber \\
   & - \underbrace{\pdr{}{r_\alpha}\pdr{}{r_{k}}\lr{\lrs{\buk\x- \buk \xtt  }H_i} }_{\bm{SST\theta }_{r_\alpha m}}
  ~.
    \label{ut-rbc_simp}
\end{align}

\textcolor{black}{The intermediate step is shown in the appendix Eqn \rf{ut-rbc_2}.} The two terms on the left-hand side are the unsteady and convective transport of scale heat flux at scale $r_\alpha$. On the right-hand side, $\bm{S_{r \alpha}} $ denotes the source term, $\bm{P\theta_{r \alpha}} $ denotes the production terms, $\bm{T\theta_{r \alpha}} $ denotes the triple correlation term, $\bm{D\theta^p_{r \alpha}} $ denotes the pressure contribution, $\bm{\epsilon\theta_{r \alpha}} $ denotes the dissipation rate, $\bm{D\theta^v_{r \alpha}} $ denotes the viscous diffusion term in physical space, $\bm{\mathcal{D}\theta_{r \alpha}} $ denotes the diffusion in scale space, $\bm{SST\theta}_{r \alpha f} $ denotes the interscale transport term due to the fluctuating velocity, and $\bm{SST\theta}_{r \alpha m} $ denotes the interscale transport term due to mean velocity.

\textcolor{black}{The energy and heat flux density function elucidates the energy and heat flux distribution among all scales of turbulence at different locations in physical space}. Terms on the right-hand side of Eqn \rf{uu46} and Eqn \rf{ut-rbc_simp} are the different mechanisms that affect the distribution of turbulent kinetic energy (heat flux) across scales. \textcolor{black}{The equations quantify the contribution of the energy and heat flux at any eddy size or scale $\mbr$ due to three types of processes;  (a) the first  due to the local production occurring at that scale due to external forces or due to mean flow field, 
(b) the second  the energy gain or loss from the processes in physical space like transport due to turbulent field, viscosity, pressure, production, dissipation, and diffusion in physical space, and 
(c) the third due to the scale space transport processes like interscale energy transport or viscous diffusion in scale space}. If we integrate Eqn \rf{uu46} and Eqn \rf{ut-rbc_simp} over all scales, $0\leq r\leq \infty$, we will get the total turbulent kinetic energy and heat flux, respectively, at a given physical location.

\begin{figure}
    \centering
    \includegraphics[width=1\linewidth]{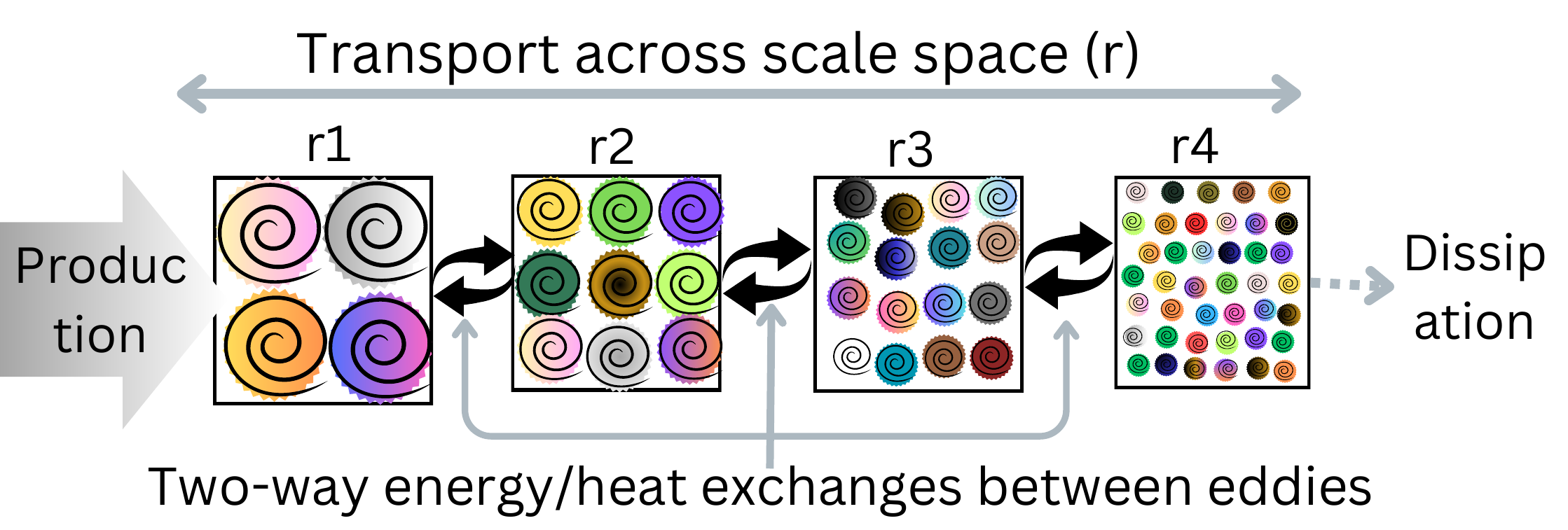}
    \caption{Schematic showing energy and heat flux transfer among eddies. The different colours of eddies indicate various orientations $\alpha$. Interscale transport depends on the direction and size of the eddies.}
    \label{fig:cascade}
\end{figure}

\textcolor{black}{ The scale space transport (SST) terms are expressed as a gradient of energy/heat flux in scale space $r$. 
These terms can have both positive and negative values at different $r_\alpha$, which means energy/heat can transfer from large scales to small scales in some direction $r_{\alpha 1}$ and from smaller scales to large scales in different directions $r_{\alpha 2}$. Hence, the energy/heat transfer cascade is a two-way process that happens in large to smaller eddies and smaller to larger eddies, as shown in the schematic in fig \ref{fig:cascade}. 
Consider a single box, depending on the position/orientation eddies of similar size/scale might have different energy/heat. In other words, energy/heat density is a function of both scale size $r$ and direction/orientation $\alpha$. The different colour mapping used in a single box signifies the possible energy/heat content difference among similar scales of turbulence at different orientations.} 


\textcolor{black}{
\citet{hill2002exact} derived a scale energy budget equation based on structure-function instead of two-point correlation used here. Some of the insights from the present analysis namely $\bm{P}_{r \alpha}, \bm{S}_{r \alpha}, \bm{D^{\nu}}_{r \alpha}$, and $\bm{D^p_{r \alpha}}$ are missing in the structure-function based scale energy equation.
Further, Marati et al.
\cite{Marati2004} modified the equations from \citet{hill2002exact} for a turbulent channel flow problem. Comparing the equations from Marati et al.\cite{Marati2004}, the proposed equations have few extra terms, mainly the production due to mean flow gradient in scale space($\frac{-1}{2}\pdr{}{r_\alpha}\lr{Q_{ik}(\bm{x},\bm{r})\pdr{\bui \xtt}{r_k}}$), the pressure-velocity correlation in scale space($\frac{1}{2\rho}\pdr{}{r_\alpha}\ol{ \pdr{u_i'\x p'\xtt}{r_i}}$), and the viscous transport term($\bm{D^{\nu}_{r \alpha}}$). The scale-space dynamics obtained from the present set of equations is more informative.   Scale-space kinetic energy transport using two-point correlations has never been attempted in buoyancy-driven flows. Togni et al.\cite{Cimarelli_Angelis_rbc_2015} studied the scale space temperature variance in RBC and examined the distribution of transport of temperature variance. Temperature variance is, however, not a direct measure of heat flux. The heat flux density transport Eqn \rf{ut-rbc_simp} has not been attempted before, as far as the authors know. }

\subsection*{Special case: Homogeneous turbulence} \label{sec:homogeneous}
In the case of homogeneous turbulence, \textcolor{black}{the statistics of any flow quantity $f$ is not varying in space, i.e., $\pdr{\av{f}}{x_i}=0$. The statistics are no longer a function of $x$, but only $r$. In other words, the flow statistics vary much more rapidly with respect to $r$ than they do with respect to $x$.} The scale space energy density (Eqn\rf{uu46}) and heat flux density transport equation (Eqn\rf{ut-rbc_simp}) can be reduced for a homogeneous case as,
\begin{align}
    & {\pdr{E(x,r_\alpha)}{t}}  = \underbrace{\frac{1}{2}\pdr{}{r_\alpha}\lr{Q_{ik}(x,r)\pdr{\bui \xtt}{r_k}}}_{\bm{P_{r_\alpha}}}
    \nonumber \\
    &  - \underbrace{\frac{1}{2}g_i \beta \pdr{}{r_\alpha}{ \lr{H_i + H_i'}}}_{\bm{S_{r_\alpha}}}
    + \underbrace{\frac{1}{2\rho}\pdr{}{r_\alpha}\lr{ \ol{ \pdr{u_i'\x p'\xtt}{r_i}} }}_{\bm{D^p_{r_\alpha}}} 
     \nonumber \\
    & + \underbrace{\frac{\nu}{2}\pdr{}{r_\alpha}\lr{ \ol{\pdr{u_i'\xtt }{x_k}\pdr{u_i'\x}{x_k} }}}_{\bm{\epsilon_{r_\alpha}}} + \underbrace{\nu{ \pdr{}{r_k}\pdr{E(x, r_\alpha)}{r_k}}}_{\mathcal{D}_{r_\alpha}} 
    \nonumber \\
    &    - \underbrace{\frac{1}{2}\pdr{}{r_\alpha}\lr{\pdr{}{r_k}\lr{Q_{ii}\lrs{\buk \x - \buk \xtt} }}}_{\bm{SST_{r\alpha,m}}} 
    \nonumber \\
    & - \underbrace{\frac{1}{2}\pdr{}{r_\alpha}\lr{\pdr{}{r_k}\lr{\av{u_i'\x u_i'\xtt \lrs{u_k'\x - u_k'\xtt}}}}}_{\bm{SST_{r\alpha, f}}} ,~\text{and}
     \label{uu46homo}
\end{align}
\begin{align}
    &  \pdr{J_i(x,r_\alpha)}{t}  = - \underbrace{\pdr{}{r_\alpha}\lr{g_i\beta \ol{T'\xtt{T'\x}}}}_{\bm{S\theta{r_\alpha}}} 
    \nonumber \\
    & + \underbrace{ \pdr{}{r_\alpha}\lr{Q_{ik}\pdr{\ol{T}\xtt}{r_{k}}}}_{\bm{P\theta_{r\alpha}}}  +  \underbrace{\frac{1}{\rho}\pdr{}{r_\alpha}\lr{\ol{ T'\xtt\pdr{ p'\x}{x_{k}}}\delta_{ik}}}_{\bm{D\theta^p_{r\alpha}}} 
    \nonumber \\
    & +   \underbrace{\pdr{}{r_\alpha}\lr{ \ol{ \pdr{T'\xtt}{x_k}{ \pdr{{u_i'\x}}{x_{k}}}}}}_{\bm{\epsilon\theta_{r\alpha}}}  +  \underbrace{\alpha {\pdr{}{r_k}\lr{\pdr{J_i(x,r_\alpha)}{r_{k}}}} }_{\bm{\mathcal{D}\theta_{r\alpha}}}
    \nonumber  \\ 
   & - \underbrace{\pdr{}{r_\alpha}\pdr{}{r_{k}}\lr{ \ol{u_i'\x  T'\xtt \lrs{u_k' \x -u_k' \xtt} }}}_{\bm{SST\theta_{r\alpha, f}}} 
   \nonumber \\
   & - \underbrace{\pdr{}{r_\alpha}\pdr{}{r_{k}}\lr{\lrs{\buk\x- \buk \xtt  }H_i} }_{\bm{SST\theta_{r\alpha, m}}}
  ~.
    \label{ut-homo}
\end{align}
We can rewrite Eqn \rf{uu46homo} and Eqn \rf{ut-homo} as,
\begin{align}
    & {\pdr{E(x,r_\alpha)}{t}}  = \bm{P}_{r\alpha} + \bm{\epsilon }_{r\alpha} + \bm{S}_{r\alpha} + \bm{SSD}_E  ,~\text{and}
    \label{reduc-E}
\end{align}
\begin{align}
    & {\pdr{J_i(x,r_\alpha)}{t}}  = \bm{P\theta_{r\alpha}} +\bm{\epsilon\theta }_{r\alpha} + \bm{S\theta_{r\alpha}} + \bm{SSD}_H .
    \label{reduc-H}
\end{align}
\textcolor{black}{Here $\bm{SSD}_E = \bm{D^p}_{r\alpha} + \bm{\mathcal{D}}_{r\alpha} + \bm{SST_{r\alpha,f}} + \bm{SST_{r\alpha,m}}$, \text{and} \\
$\bm{SSD}_H =  \bm{D^p\theta_{r\alpha}} + \bm{\mathcal{D}\theta_{r\alpha}} + \bm{SST \theta}_{r\alpha,f} + \bm{SST \theta}_{r\alpha,m} $ are the interscale transport terms. Without the interscale transport terms, Eqn \rf{reduc-E} and Eqn \rf{reduc-H} become the familiar form as ,}
\begin{align}
    & {\pdr{E(x,r_\alpha)}{t}}  = \bm{P}_{r\alpha} +\bm{\epsilon }_{r\alpha} + \bm{S}_{r\alpha},~\text{and}
    \label{reduc-E2}
\end{align}
\begin{align}
    & {\pdr{J_i(x,r_\alpha)}{t}}  = \bm{P\theta_{r\alpha}} +\bm{\epsilon\theta }_{r\alpha} + \bm{S\theta_{r\alpha}} .
    \label{reduc-H2}
\end{align}
This is the single-point energy budget equation for homogeneous isotropic turbulence, which dictates that the time rate of energy (heat transfer) equal the production(source) and the dissipation rate. \textcolor{black}{We demonstrate the contribution of the production, dissipation, interscale transport and source terms in scale-space using a case study, namely Rayleigh-Benard convection. }

\textcolor{black}{\section{Case study: 2D Rayleigh-Benard Convection}\label{sec:RBC}}

\textcolor{black}{The 2D Rayleigh-Benard convection(RBC) is a canonical set-up for studying bouyancy driven flows. The literature on various aspects of RBC is vast \cite{benard1900etude, rayleigh1916lix, kadanoff2001turbulent, ahlers2009turbulent, VERZICCO_CAMUSSI_2003, Weiss2023, heat_transport}; here, we focus on the scale-space transport in turbulent RBC. Recently, \citet{Zhou2024} studied the effect of low Prandtl numbers on global and local statistics of turbulent flow using a 2D RBC case. Similarly, \citet{zhao_cb_2024} conducted a study of droplet transmission in a poorly ventilated, where the major driver of the flow is the human body heat. Here, for simplicity without losing generality, a 2D RBC data set is used to demonstrate the new inferences that can be deduced from the proposed equations. The details of the data set and numerical set-up are available in an earlier publication \cite{Sharma2023} demonstrating the suitability of entropically damped artificial compressibility method for buoyancy-driven flows. The essentials of the method, namely the governing equations and non-dimensionalization, are discussed below.}

\textcolor{black}{Retaining the same variables for non-dimensional quantities, the governing non-dimensional equations are:
\begin{align}
    \pdr{ u_k}{x_k} = 0  \label{rbc1},
\end{align}
\begin{align}
    \pdr{u_i}{t} + u_k \pdr{u_i}{x_k} = -\pdr{p}{x_k}\delta_{ik} + \frac{1}{Re}\pdr{^2u_i}{x_k^2} + PrEc\theta {g}_i, 
    \label{rbc2}
\end{align}
\begin{align}
\pdr{\theta}{t} + u_k\pdr{\theta}{x_k} = \frac{1}{\sqrt{PrRa}}\pdr{}{x_k}\pdr{\theta}{x_k}.
\label{rbc3}
\end{align}
Here, $u_i$ is the component of the velocity vector, $p$ is the pressure and $\theta$ is the non-dimensional temperature. $\delta_{ik}$ is the Kronecker delta, and ${g}_i$ represent the gravity which has the components $(0,-1,0)$. The characteristic scale used for non-dimensionalization are: the cell height $H$ as a length scale, the free fall speed $u_0=\sqrt{g\beta\Delta T H}$ as velocity scale \cite{VERZICCO_CAMUSSI_2003}, free fall time $t_0=H/u_0$ as a time scale, and $\rho u_o^2$ for pressure. The dimensionless temperature is defined as \cite{clausen2013entropically},
 \begin{equation}
\theta=\frac{(T-T_0)C_{p_{0}}}{(u_0^2\text{ Pr})}= \frac{(T-T_0)}{\Delta T}\frac{1}{\text{Pr }\text{Ec}}.
\label{theta}
  \end{equation}
The above scaling gives the non-dimensional numbers as; Eckert number $Ec=u_0^2/\Delta T C_{p_{0}}$, Prandtl number $Pr = \nu /\alpha$, Rayleigh number $Ra = g\beta \Delta T H^3 / \nu \alpha$, and Reynolds number $Re = \sqrt{Ra/Pr}$. The simulations are done for $Pr=1/Ec=0.7$, which reduces Eqn \rf{rbc2} to the familiar form in the RBC community \cite{VERZICCO_CAMUSSI_2003, stevens2018turbulent, SHISHKINA_THESS_2009, schumacher2018transition}.
Instead of a pressure-Poisson equation, the pressure is solved using the Entropically Damped Artificial Compressibility (EDAC) equation, which is of the form,
 \begin{equation}   
      \frac{Dp}{Dt} = -\frac{1}{M^2} \nabla \cdot \mathbf{V} + \frac{\gamma}{Re Pr} \nabla^2  p.
      \label{edac}
 \end{equation}
Derivation and detailed explanations about EDAC can be found in \citet{clausen2013entropically,Sharma2023, kajzer2020weakly}. 
}

\textcolor{black}{\subsection{Data set}}
\textcolor{black}{\subsection*{Numerical setup}
\begin{figure}
    \centering
    \includegraphics[width=1\linewidth]{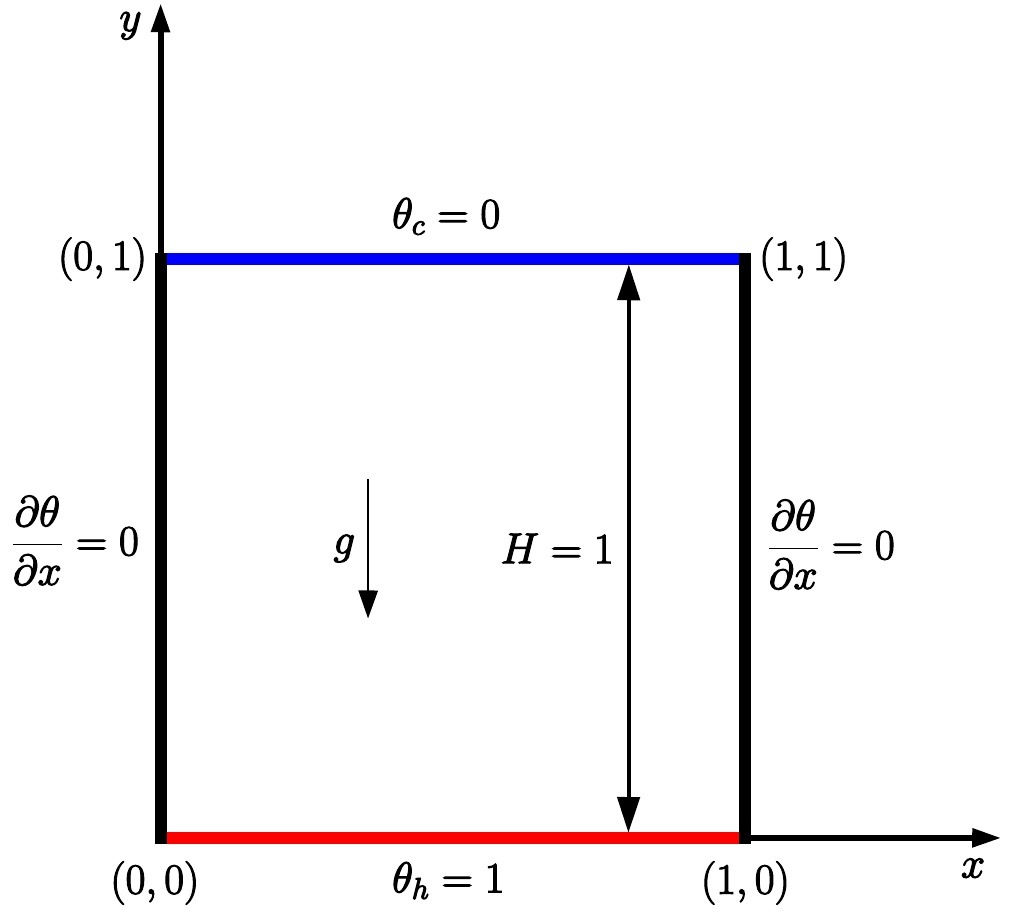}
    \caption{\textcolor{black}{Computational domain and thermal boundary conditions of the RBC. All the boundaries are no-slip walls. The subscripts $h$ and $c$ denote hot and cold walls, respectively.}}
    \label{fig:domain_RBC}
\end{figure}
}

\textcolor{black}{
The governing equations \rf{rbc1}-\rf{rbc3} are solved using the EDAC method for two-dimensional RBC to obtain the velocity, temperature, and pressure fields.  A schematic diagram of the computational domain and the boundary conditions for the RBC is shown in fig.\ref{fig:domain_RBC}. The domain is a square cavity of unit length. The top wall of the domain is kept at a constant non-dimensional temperature of $\theta_c = 0$, and the bottom wall is maintained at a constant non-dimensional temperature of $\theta_h = 1$. The side walls are adiabatic, and a Neumann boundary condition ($\partial \theta / \partial x$) is applied to these walls. All the walls are no-slip boundaries. The RBC flow simulated for Rayleigh numbers of $10^7$ and $10^8$ are shown here.  The simulation is carried out using an in-house code with $6^{th}$-order compact difference scheme, and EDAC equation for solving pressure. The time stepping is done using the classical Runge-Kutta method. More details of the scheme can be found in \citet{Sharma2023}, including the validation and grid sensitivity studies. The grid sensitivity and validation are not repeated here for brevity.}

\textcolor{black}{\subsection*{Mean and fluctuating fields}}
 \begin{figure}
    \centering
    \includegraphics[width=1\linewidth,]{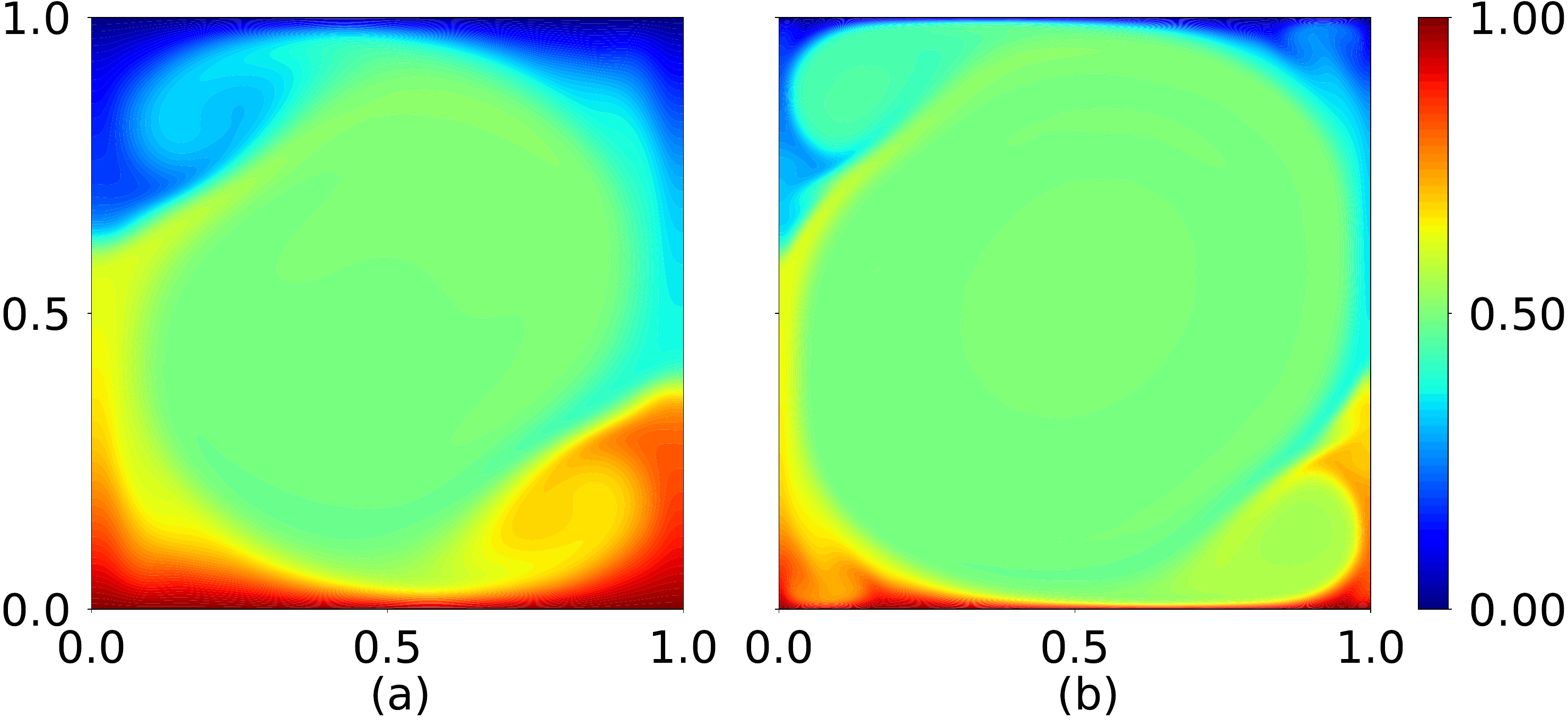}
    \caption{\textcolor{black}{Time-averaged temperature($\overline{\theta}$) contours for RBC flows for various Rayleigh numbers:(a) $10^7$, and (b) $10^8$.}}
    \label{fig:mean_temp}
\end{figure}

 \textcolor{black}{The data set from simulations for  RBC for Rayleigh numbers $10^7$ and  $10^8$ is used in the present study to evaluate the scale-space transport equations. The statistical samples are accumulated once the flow achieves a statistically steady state. The samples are then used to take Reynolds averaging to get the mean flow field. The mean temperature field is shown in fig.\ref{fig:mean_temp}. The mean temperature contour shows the major features of the flow field, the Large Scale Circulation(LSC) and the two re-circulating eddies on the top left and bottom right corners. }

\begin{figure}
    \centering
    \includegraphics[width=1\linewidth,]{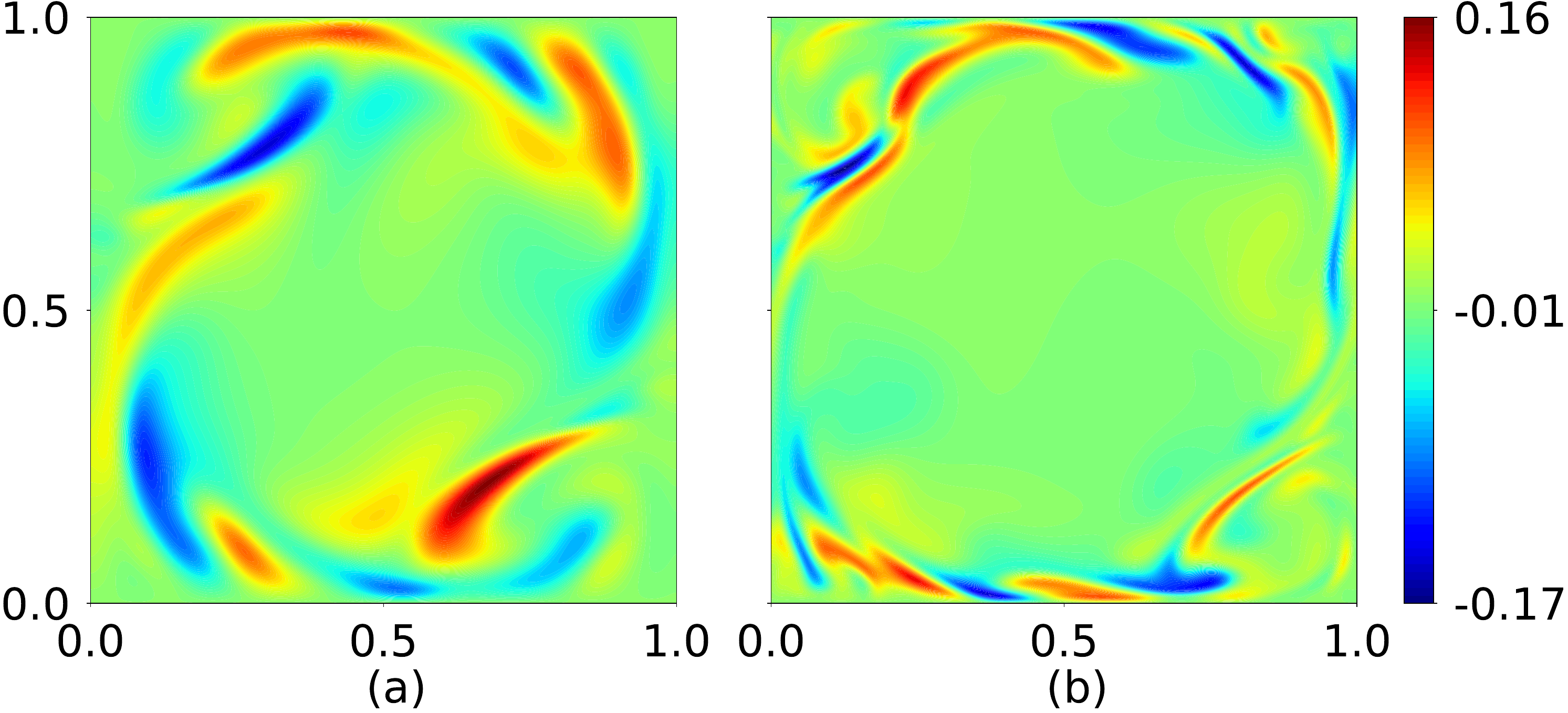}
    \caption{\textcolor{black}{Fluctuating temperature($\theta'$) contours for RBC flows for  (a) $Ra=10^7$, and (b) $Ra=10^8$.}}
    \label{fig:fluc_temp}
\end{figure}

 \textcolor{black}{The fluctuating quantities are calculated by subtracting the mean from the instantaneous values. The fluctuating temperature field for Ra $10^7$, and $10^8$ cases are shown in fig.\ref{fig:fluc_temp}. The turbulent fluctuations are mostly along the periphery of the LSC, and the recirculating eddies for these two Rayleigh numbers of RBC \cite{VERZICCO_CAMUSSI_2003}.}

\textcolor{black}{\subsection*{Single-point statistics}}

\begin{figure}
    \centering
    \includegraphics[width=1\linewidth,]{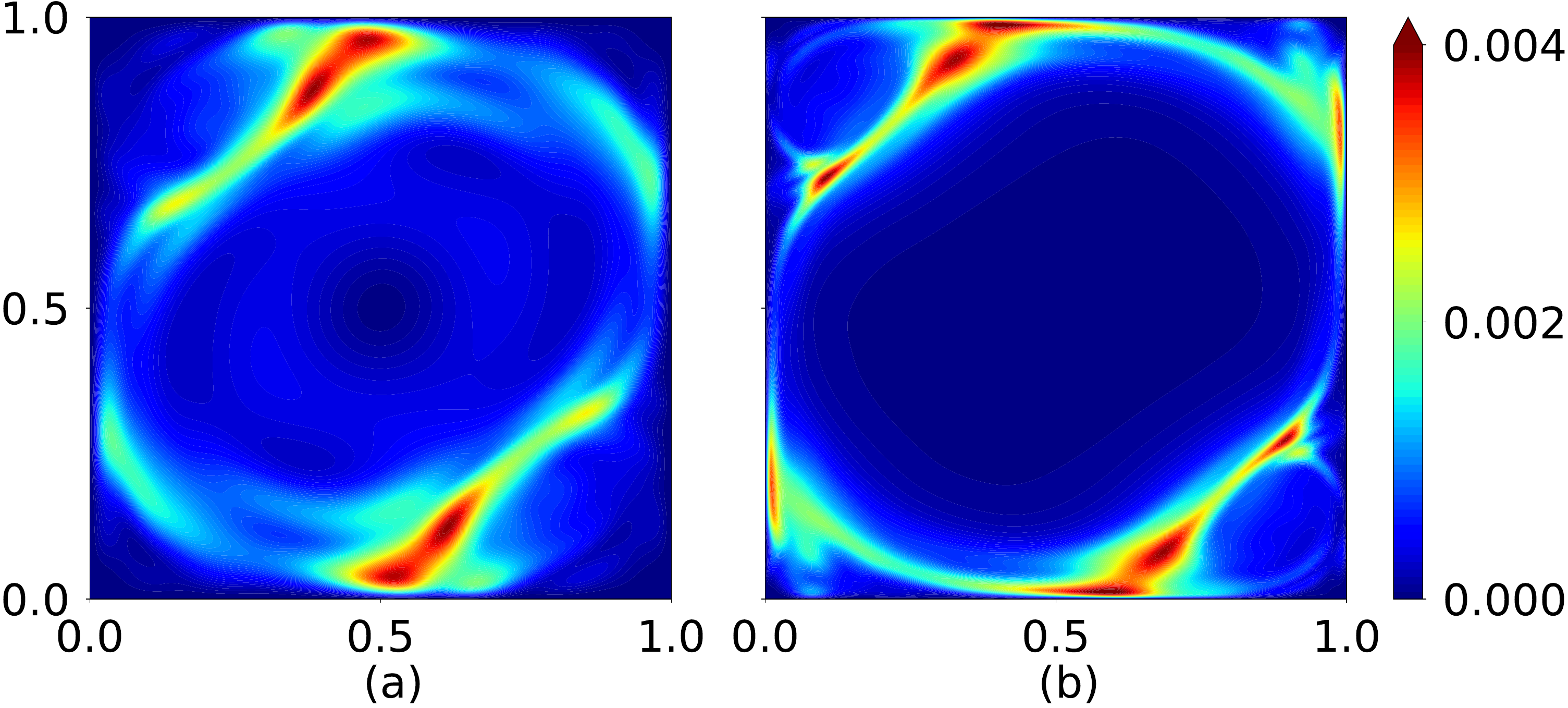}
    \caption{\textcolor{black}{Turbulent kinetic energy $\frac{1}{2}\overline{u_i' u_i'}$ contours  for Rayleigh numbers (a) $10^7$, and (b) $10^8$.}}
    \label{fig:tke}
\end{figure}
\begin{figure}
    \centering
    \includegraphics[width=1\linewidth,]{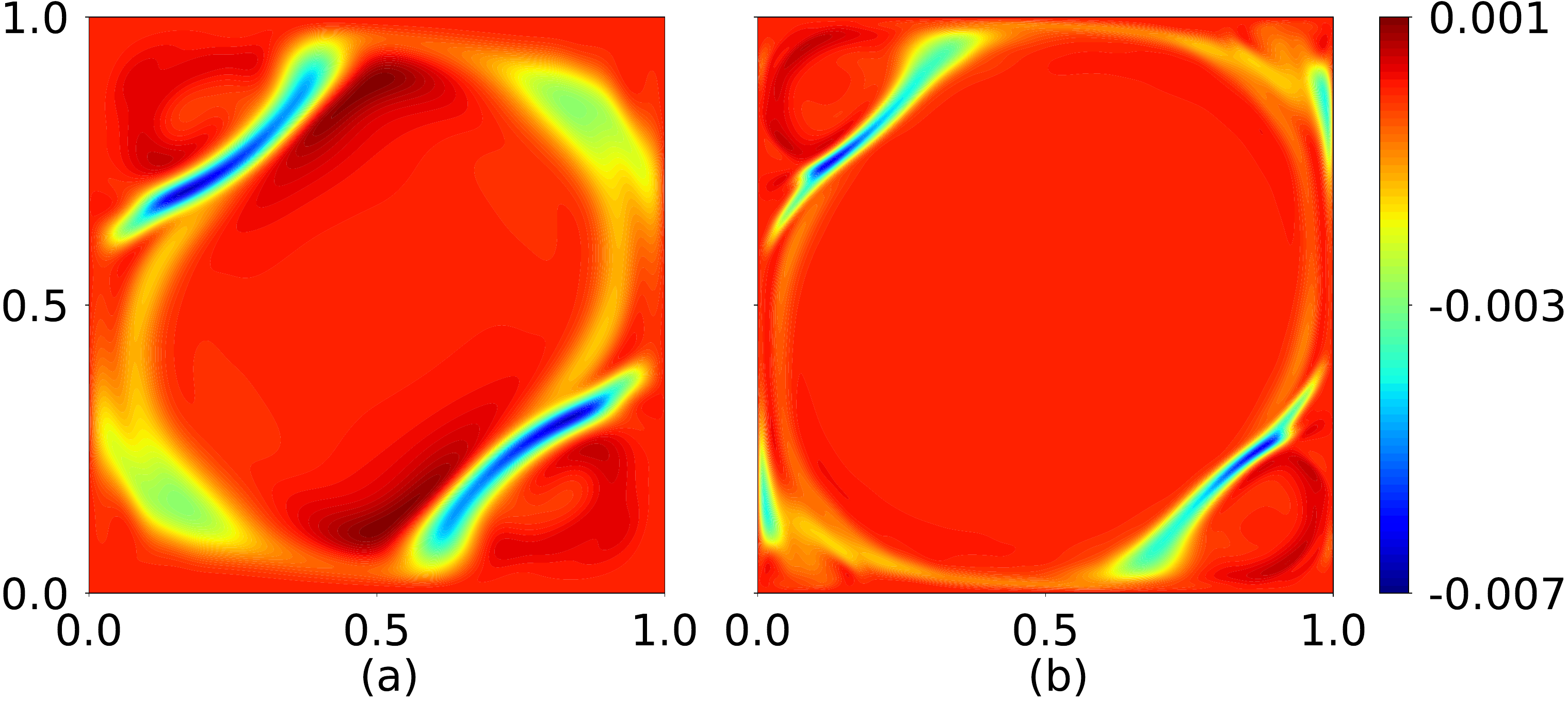}
    \caption{\textcolor{black}{Contours of turbulent heat flux in y-direction ($\overline{\text{v}'\theta'}$)  for Rayleigh numbers: (a) $10^7$, and (b) $10^8$.}}
    \label{fig:THF}
\end{figure}

\textcolor{black}{It is customary to corroborate the turbulence in a flow with the computation of turbulent kinetic energy(TKE). Single-point statistics quantities of TKE and turbulent heat flux(THF) are calculated for $Ra=10^7$ and $10^8$. The contours of turbulent kinetic energy(TKE) defined earlier as $\frac{1}{2}\overline{u'_i(\mbx) u'_i(\mbx)}$ are shown in fig.\ref{fig:tke}. The peak of TKE happens near the wall close to the point where LSC and the recirculating eddies interact. A similar pattern has been observed in many literature\cite{DELORTLAVAL2022122453, PALLARES2002346, heat_transport, Weiss2023} where the kinetic energy is zero at the boundaries, reached maximum value near the boundary at approximately thermal boundary layer thickness distance from the wall, and flattened at the center region.}

\textcolor{black}{Similarly, turbulent heat flux is also calculated for both cases. The turbulent heat flux(THF) in y direction ($\overline{\text{v}'(\mbx)\theta'(\mbx)}$) is shown in fig.\ref{fig:THF}, where $\text{v}'(\mbx)$ is the fluctuating velocity component in the y-direction. The correlation is positive for the majority of the domain except near the periphery of the LSC. This implies that when $\text{v}'(\mbx)$ is positive, the fluid carries hot fluid ($\theta'(\mbx)> 0$) upwards and vice versa for most parts of the domain.}

\textcolor{black}{\subsection{Energy and heat flux density in scale-space}}

\textcolor{black}{
The data set discussed in the previous section is used to compute the scale space quantities. 
The transport equations for energy density and heat flux density in a dimensionless form can be formed from Eqn \rf{uu46} and Eqn \rf{ut-rbc_simp} using the reference length and velocity scale for RBC mentioned above. Similarly, the transport equations for two-point correlation in the non-dimensional form can be obtained from Eqn \rf{uu44_ii} and Eqn \rf{ut-rbc}, which is shown in Appendix Eqn \rf{Qii-rbc_nd} and Eqn \rf{ut-rbc_nd}. In the following discussion, only the terms from transport equations for energy density and heat flux density are computed.}

\textcolor{black}{
The energy density and heat flux density equations in the dimensionless form are,
\begin{align}
    & {\pdr{E(\bm{x},r_\alpha)}{t}} + {\buk \x \pdr{E(\bm{x,}r_\alpha)}{x_k}} =  \underbrace{\frac{1}{2}\pdr{}{r_\alpha}\lr{\pdr{T_{iik}(\bm{x},\bm{r})}{x_k}}}_{\bm{T_{r_\alpha}}} 
    \nonumber \\
    &+ \underbrace{  {\frac{1}{2}\pdr{}{r_\alpha}\lr{ Q_{ki}(\bm{x},\bm{r})}\pdr{\bui \x }{x_k}}  + \frac{1}{2}\pdr{}{r_\alpha}\lr{Q_{ik}(\bm{x},\bm{r})\pdr{\bui \xtt}{r_k}}}_{\bm{P_{r_\alpha}}}
    \nonumber \\
    &  - \underbrace{\frac{1}{2} PrEc\pdr{}{r_\alpha}{ \lr{H_i(\bm{x},\bm{r}) + H_i'(\bm{x},\bm{r})}} g_i}_{\bm{S_{r_\alpha}}}  
    \nonumber \\
    & - \underbrace{\frac{1}{2Re}\pdr{}{r_\alpha}\lr{\pdr{}{x_k}\lr{\ol{u_i'\xtt \pdr{u_i'\x}{x_k}}}}}_{\bm{D^v_{r_\alpha}}}
    \nonumber \\
    & + \underbrace{\frac{1}{2}\pdr{}{r_\alpha}\lr{\ol{ \pdr{u_i'\xtt p'\x}{x_i}} + \ol{ \pdr{u_i'\x p'\xtt}{r_i}} }}_{\bm{D^p_{r \alpha}}} 
     \nonumber \\
    & + \underbrace{\frac{1}{2Re}\pdr{}{r_\alpha}\lr{ \ol{\pdr{u_i'\xtt }{x_k}\pdr{u_i'\x}{x_k} }}}_{\bm{\epsilon_{r_\alpha}}} + \underbrace{\frac{1}{Re}{ \pdr{}{r_k}\pdr{E(\bm{x}, r_\alpha)}{r_k}}}_{\mathcal{D}_{r\alpha}} 
    \nonumber \\
    &    - \underbrace{\frac{1}{2}\pdr{}{r_\alpha}\lr{\pdr{}{r_k}\lr{Q_{ii}(\bm{x},\bm{r})\lrs{\buk \x - \buk \xtt} }}}_{\bm{SST_{r \alpha,m}}} 
    \nonumber \\
    & - \underbrace{\frac{1}{2}\pdr{}{r_\alpha}\lr{\pdr{}{r_k}\lr{\av{u_i'\x u_i'\xtt \lrs{u_k'\x - u_k'\xtt}}}}}_{\bm{SST_{r \alpha, f}}} ~,
     \label{eq:Era_rbc}
\end{align}
\begin{align}
    &  \pdr{J_i(x,r_\alpha)}{t} +  {\buk \x \pdr{J_i(x,r_\alpha)}{x_{k}}} = - \underbrace{\pdr{}{r_\alpha}\lr{PrEc\ol{\theta'\xtt{\theta'\x}}} g_i}_{\bm{S\theta_{r_\alpha}}} 
    \nonumber \\
    & \underbrace{ + \pdr{H_k}{r_\alpha}{\pdr{\bui\x}{x_{k}}}  + \pdr{}{r_\alpha}\lr{Q_{ik}\pdr{\ol{\theta}\xtt}{r_{k}}}}_{\bm{P\theta_{r\alpha}}} +  \underbrace{\pdr{}{r_\alpha}{\pdr{\mathcal{T}_{ik}}{x_k}}}_{T\theta_{r\alpha}}
    \nonumber \\
    &  +  \underbrace{\pdr{}{r_\alpha}\lr{\ol{ \theta'\xtt\pdr{ p'\x}{x_{k}}}\delta_{ik}}}_{\bm{D\theta^p_{r\alpha}}}  \nonumber \\
    &-  \underbrace{\frac{1}{Re} \pdr{}{r_\alpha}\lr{ \ol{ \pdr{}{x_k}\lr{\theta'\xtt \pdr{{u_i'\x}}{x_{k}}}}}}_{\bm{D\theta^v_{r\alpha}}} \nonumber \\
     &+   \underbrace{\frac{1}{Re}\pdr{}{r_\alpha}\lr{ \ol{ \pdr{\theta'\xtt}{x_k}{ \pdr{{u_i'\x}}{x_{k}}}}}}_{\bm{\epsilon\theta_{r\alpha}}} +  \underbrace{\frac{1}{\sqrt{PrRa}} {\pdr{}{r_k}\lr{\pdr{J(x,r_\alpha)}{r_{k}}}} }_{\bm{\mathcal{D}\theta_{r\alpha}}}
    \nonumber  \\ 
   & - \underbrace{\pdr{}{r_\alpha}\pdr{}{r_{k}}\lr{ \ol{u_i'\x  \theta'\xtt \lrs{u_k' \x -u_k' \xtt} }}}_{\bm{SST\theta}_{r\alpha, f}} 
   \nonumber \\
   & - \underbrace{\pdr{}{r_\alpha}\pdr{}{r_{k}}\lr{\lrs{\buk\x- \buk \xtt  }H_i} }_{\bm{SST\theta }_{r\alpha, m}}
  ~.
    \label{eq:J-rbc_simp}
\end{align}}

\textcolor{black}{The notations in the underbraces have the same definitions as the dimensional equations Eqn \rf{uu46} and Eqn \rf{ut-rbc_simp}. The physical meaning of these terms is discussed in the previous section. Here, interpretation is carried out using data from the simulation of RBC.}

\textcolor{black}{To calculate the two-point correlations and the terms in the energy/heat flux density transport equation, we fix the first point ($x_1$) at the middle of the domain. This is a representative point, and this is selected because, for the discussion of single point statistics of turbulent RBC, the center point is often chosen. 
The second point ($x_2$) is then varied throughout the domain to calculate two-point velocity, velocity-temperature, and other correlations. 
These correlations are then used to calculate the terms of the equations Eqn \rf{eq:Era_rbc} and Eqn \rf{eq:J-rbc_simp}. For the current analysis, we limit ourselves to four major terms from these equations: production($\bm{P_{r \alpha}}$), dissipation($\bm{\epsilon}_{r \alpha}$), scale space transport($\bm{SST}_{r \alpha,m}$,$\bm{SST}_{r \alpha,f}$), and buoyancy source term($\bm{S}_{r \alpha}$). The corresponding terms from heat flux density equation are $\bm{P\theta_{r \alpha}},\bm{\epsilon\theta}_{r \alpha} ,\bm{SST\theta}_{r \alpha,m}, \bm{SST\theta}_{r \alpha,f}$ and $\bm{S\theta}_{r \alpha}$.}

\textcolor{black}{\subsection*{Energy density transport}}

\textcolor{black}{ In this section the terms from the energy density transport equation (Eqn \rf{eq:Era_rbc}) are examined, and followed by the terms from heat flux density transport equation (Eqn \rf{eq:J-rbc_simp}) in the next section. Henceforth, all the contours are plotted in the scale space of the $r_x$-$r_y$ plane, with the origin being the fixed $x_1$ point, which is at the middle of the domain.}

\begin{figure}
    \includegraphics[width=1\linewidth,]{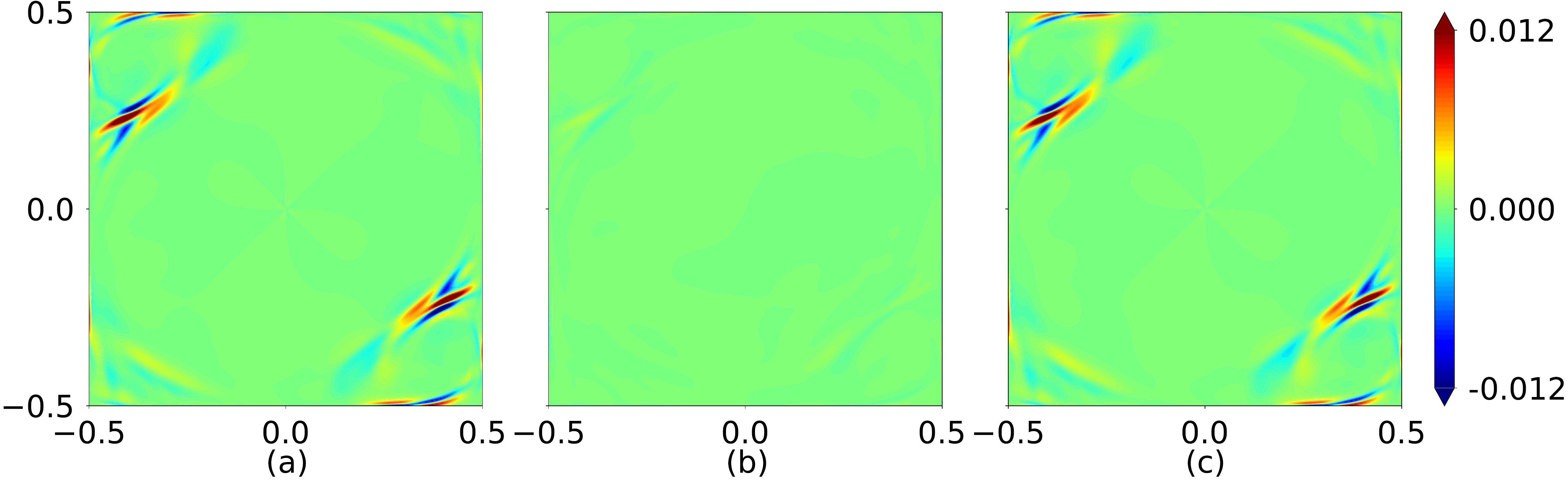}
    \caption{\label{fig:prod_ra108}\textcolor{black}{Contours in $r_x$-$r_y$ plane of (a) total production of energy density ($\bm{P}_{r \alpha}$), (b) production due to mean flow gradient in physical space, and (c) production due to mean flow gradient in scale space for $Ra = 10^8$.}}
\end{figure}



\textcolor{black}{The production term, $\bm{P}_{r\alpha}$ is split into two parts: $\frac{1}{2}\pdr{}{r_\alpha}\lr{ Q_{ki}(\bm{x},\bm{r})}\pdr{\bui \x }{x_k}$ due to the mean flow gradient in physical space and $\frac{1}{2}\pdr{}{r_\alpha}\lr{Q_{ik}(\bm{x},\bm{r})\pdr{\bui \xtt}{r_k}}$ due to the gradient in scale space. The fig.\ref{fig:prod_ra108} shows the contours of total production of energy density from mean flow, production due to mean flow gradient in physical space, production due to mean flow gradient in scale space for $Ra = 10^8$. The production due to the mean flow gradient in physical space contributes less towards the production terms as it only depends on the mean velocity gradient at the location $\mbx$. This part is highest at a scale corresponding to the inflection point of the two-point correlation. Meanwhile, the second term, production due to the mean flow gradient in scale space, depends on the choice of $r$ and direction, $\alpha$. These terms are highest in the regions where mean velocity gradients are large. This term contributes the most towards the total production. As seen in fig.\ref{fig:prod_ra108}a-c, there are negative production regions where the energy flow is from turbulent fluctuations to mean flow at some $r$ and $\alpha$. The negative production can be identified in literature\cite{Cimarelli2019}, and for RBC flows it was observed by Tsinober et al.\cite{RBC_negTsinober2003}. The novel outcome of this analysis is the contribution of the second term to production due to the mean flow gradient in scale space for inhomogeneous flow, which is inaccessible without Eqn \rf{eq:Era_rbc}.}

\begin{figure}
    \centering
    \includegraphics[width=1\linewidth,]{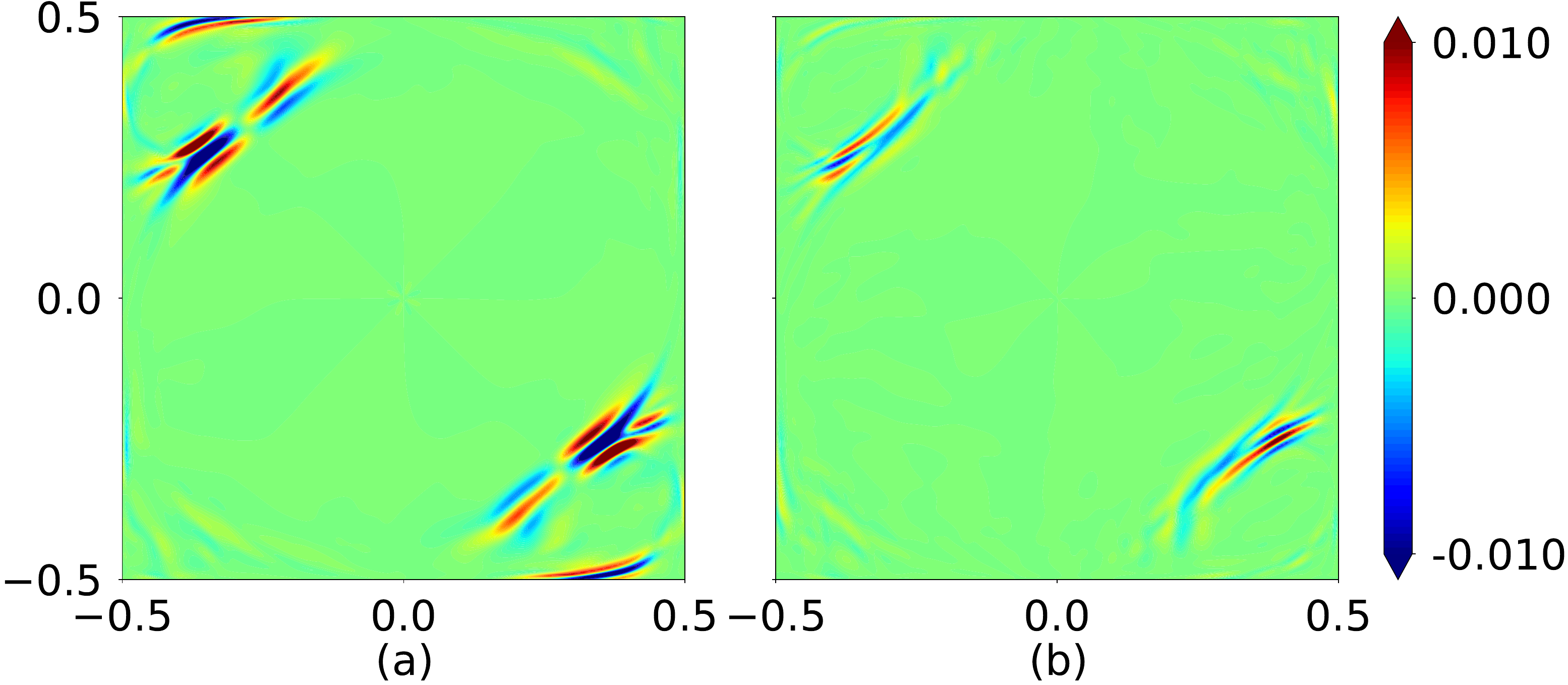}
    \caption{\textcolor{black}{Contours of  interscale transport of energy density (a) due to mean flow field ($\bm{SST}_{r \alpha, m}$), and (b)  due to fluctuating flow field ($\bm{SST}_{r \alpha, f}$) in RBC flows for $Ra = 10^8$.}}
    \label{fig:SST_mean}
\end{figure}


\textcolor{black}{Another important feature of these equations is the ability to calculate energy transfer in scale space or interscale energy transport. The interscale transport term has two parts, one caused by the inhomogeneity in mean velocity ($\bm{SST}_{r \alpha,m}$) and the other due to the inhomogeneity in fluctuating velocity($\bm{SST}_{r \alpha,f}$). The former is absent for a homogeneous flow, while the latter is the common mechanism for energy transfer in homogenous turbulence, which also appears in wavenumber space \cite{Lesieur2008}. The contours of these terms are shown in fig.\ref{fig:SST_mean}. The transport due to mean flow inhomogeneity is the major contributor to the scale space energy transport. There are scales with positive and negative scale space energy transfers, showing both forward and inverse energy transfer. The inhomogeneous nature of the flow field makes the energy transport depend on the orientation $\alpha$ of the separation vector. The contribution of mean flow inhomogeneity to the interscale transport using structure function was reported previously \cite{hill2002exact,Marati2004}, which corroborates the present budget equation.}

\begin{figure}
    \centering
    \includegraphics[width=1\linewidth,]{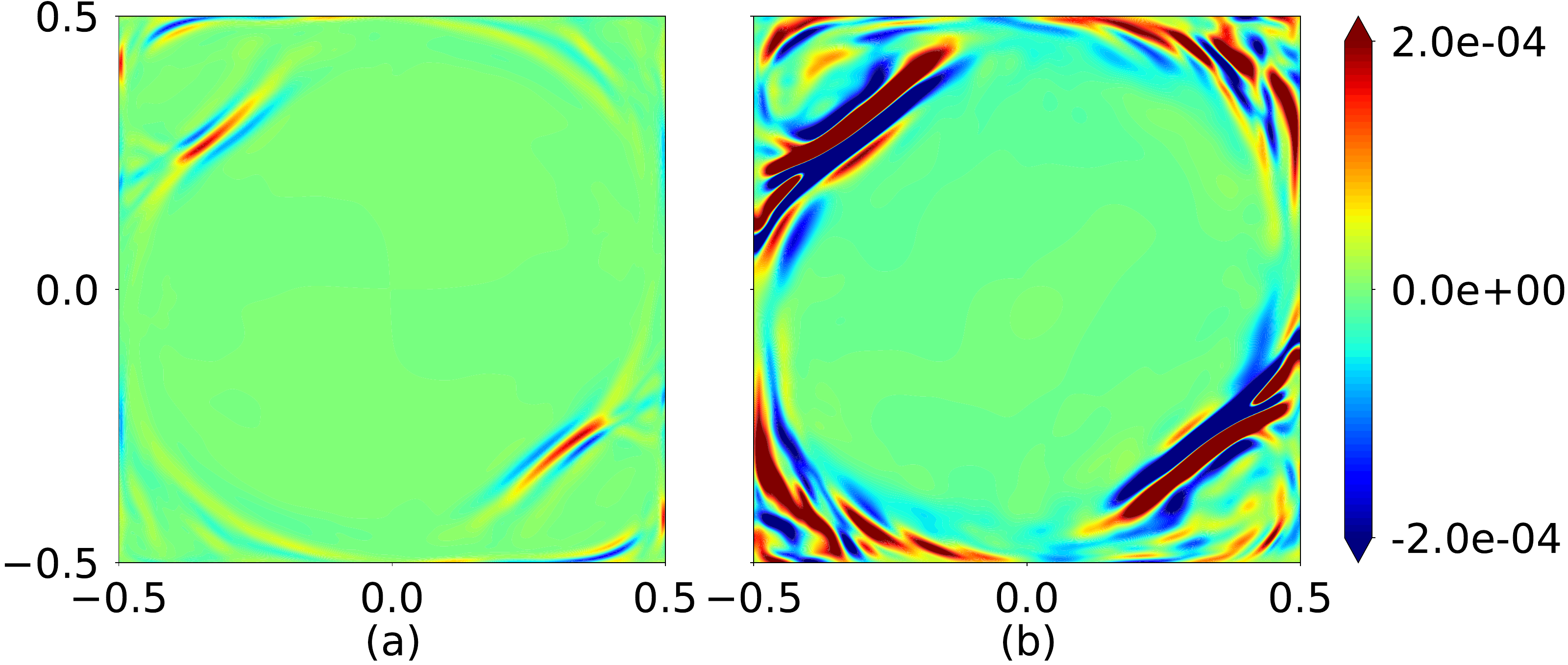}
    \caption{\textcolor{black}{Contours of (a) dissipation of energy density ($\bm{\epsilon}_{r \alpha}$), and (b) production of energy density due to buoyancy($\bm{S}_{r \alpha}$) for $Ra = 10^8$.}}
    \label{fig:dissi_buoy_ra108}
\end{figure}

\textcolor{black}{The contours of viscous dissipation ($\bm{\epsilon}_{r \alpha}$) and the production of energy density due to buoyancy ($\bm{S}_{r \alpha}$) are shown in fig.\ref{fig:dissi_buoy_ra108}. The dissipation at any scale depends on the two-point velocity gradient correlation at the corresponding scale separation and orientation, which is shown in fig.\ref{fig:dissi_buoy_ra108}a. Since the equations are local budget terms, dissipation can be locally positive. The peak of the dissipation happens at the scale corresponding to the maximum value of the derivative of the two-point gradient correlation. The bulk of the dissipation happens on scales near the edge of LSC and the regions near the thermal boundary layer\cite{Zhang_Zhou_Sun_2017}. The fig.\ref{fig:dissi_buoy_ra108}b shows the contour of production of energy density due to buoyancy(source). The buoyancy production depends on the sum of the two-point velocity at $x_1$-temperature at $x_2$ and temperature at $x_1$-velocity at $x_2$ correlation. The peak value happens at a scale corresponding to the peaks of the gradient of these correlations. Evidently, at $Ra=10^8$  production of energy density (positive and negative) is more along the periphery of LSC.}

\textcolor{black}{ \subsection*{Heat flux density transport}}
\textcolor{black}{The terms from the heat flux density transport Eqn \rf{eq:J-rbc_simp} are examined in this section for  $Ra = 10^8$. Similar to the energy density transport equation,  the first point $x_1$ is fixed, and the second point $x_2$ is varied across the domain to calculate the terms from Eqn \rf{eq:J-rbc_simp}. Unlike energy density, heat flux density is a vector, and for brevity, we limit our analysis to only the y-direction (gravity direction) heat flux density.} 

\begin{figure}
    \centering
    \includegraphics[width=1\linewidth,]{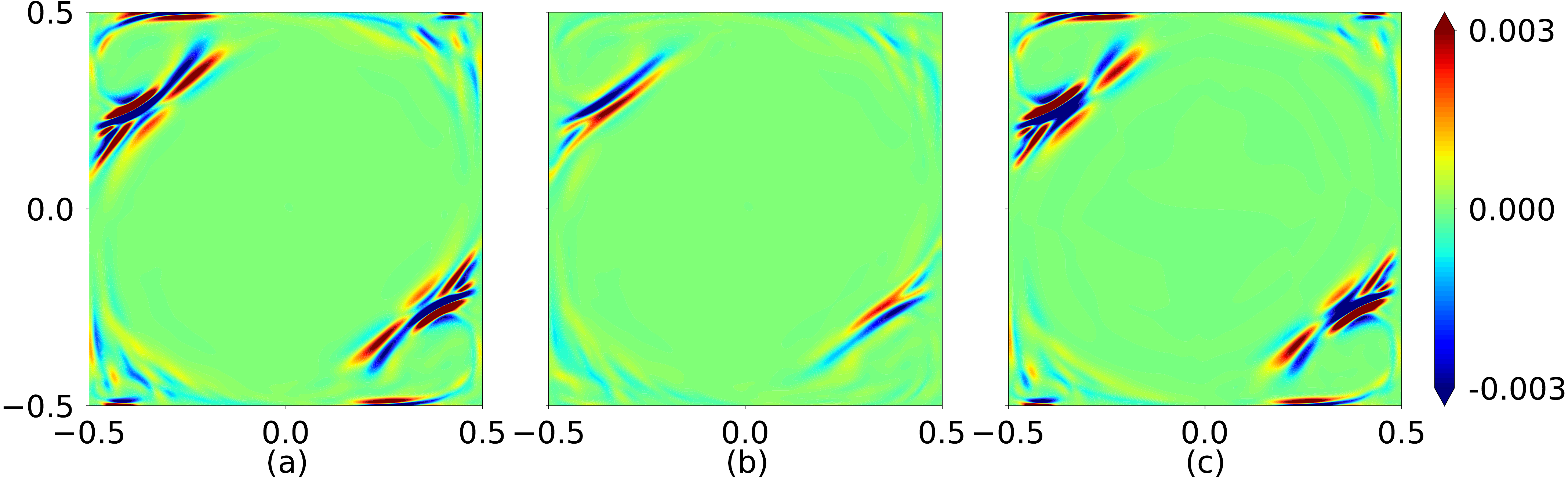}
    \caption{\textcolor{black}{Contours of (a) total heat flux density production ($\bm{P\theta}_{r \alpha}$), (b) production due to mean velocity gradient in physical space, and (c) production due to mean temperature gradient in scale space for $Ra = 10^8$.}}
    \label{fig:prod_hfy_ra108}
\end{figure}

\textcolor{black}{
The contours of total heat flux density production ($\bm{P\theta}_{r \alpha}$), production due to mean flow gradient in physical space and production due to mean flow gradient in scale space are shown in fig.\ref{fig:prod_hfy_ra108}. The production due to the mean flow gradient in physical space is the first part of the total production term $\bm{P\theta}_{r \alpha}$ in Eqn \rf{eq:J-rbc_simp}, which depends on the mean velocity gradient at the location $\mbx$. The highest value of this term happens at a scale that corresponds to the inflection point of the two-point temperature-velocity correlation. The second part of the production term is due to the mean temperature gradient in scale space. The second term contributes the most to the production of heat flux density in the y-direction. Here also, the equation captures negative production on some scales. }

\begin{figure}
    \centering
    \includegraphics[width=1\linewidth,]{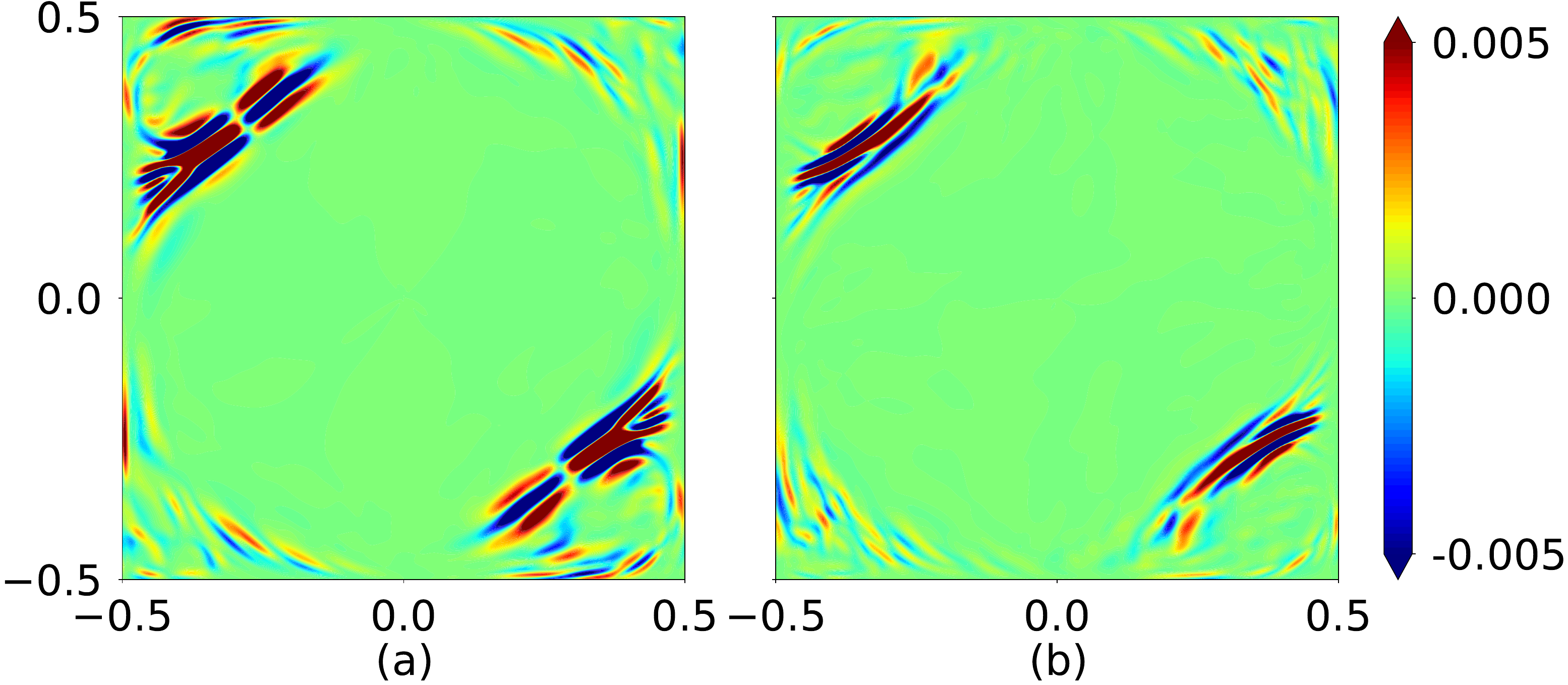}
    \caption{\textcolor{black}{Contours of  scale space transport of heat flux density (a) due to the mean field ($\bm{SST\theta}_{r\alpha,m}$), and (b) due to the fluctuating field ($\bm{SST\theta}_{r\alpha,f}$)  for $Ra = 10^8$.}}
    \label{fig:SST_mean_hfy}
\end{figure}

\textcolor{black}{Fig.\ref{fig:SST_mean_hfy} shows contours of the interscale space transport due to mean ($\bm{SST \theta}_{r \alpha, m}$) and fluctuating flow field ($\bm{SST \theta}_{r \alpha, f}$). The scale space transport terms give the net transport of energy at any scale r. The term due to the mean flow arises from the inhomogeneity in mean flow. This term vanishes for homogenous flows. The second term is due to the difference in fluctuating field at the two points. This is the canonical energy transport mechanism in homogeneous turbulence. These two terms contribute equally to the scale space transport at all scales. The scale space transport terms are positive and negative, showing both the forward and inverse cascading processes. The intensity and the range of scales depend on the orientation of the scale $\mbr$ and the position $\mbx$. The contours show that interscale transfer is largest near the vicinity of small circulation at the corners near hot and cold walls. This is an intuitively known fact \cite{kadanoff2001turbulent}, which can be quantified with the present analysis.}

\begin{figure}
    \centering
    \includegraphics[width=1\linewidth,]{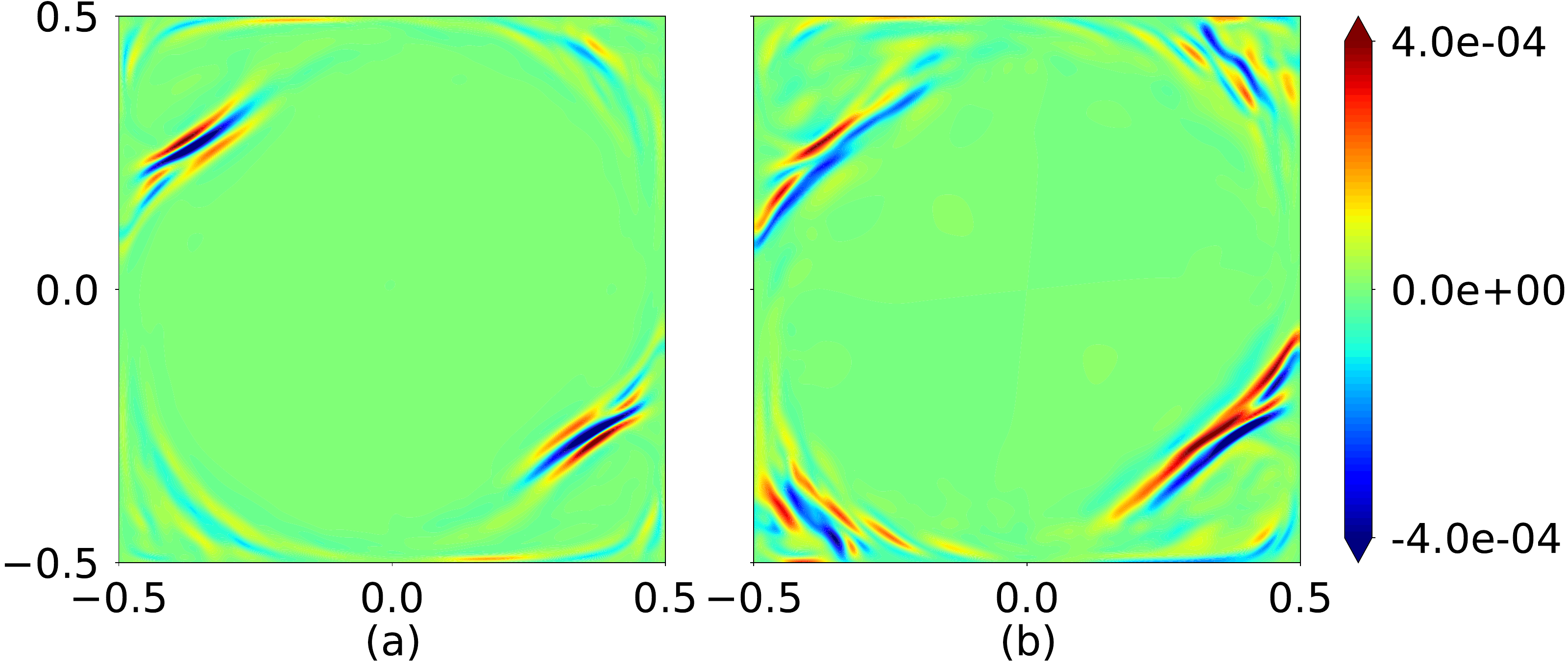}
    \caption{\textcolor{black}{Contours of (a) dissipation of heat flux density ($\bm{\epsilon\theta}_{r\alpha}$), and (b) production of heat flux density due to buoyancy ($\bm{S\theta}_{r\alpha}$) for $Ra = 10^8$.}}
    \label{fig:reamin_hfy_ra108}
\end{figure}

\textcolor{black}{The contours of viscous dissipation($\bm{\epsilon \theta}_{r \alpha}$) and production of heat flux density due to buoyancy($\bm{S\theta}_{r \alpha}$) are shown in fig.\ref{fig:reamin_hfy_ra108}. Fig.\ref{fig:reamin_hfy_ra108}a shows the contours of viscous dissipation of the heat flux density, which depend on the two-point correlation of temperature and velocity gradients. The peak of the term occurs at a scale that corresponds to the maximum of the two-point gradient correlation. The production due to buoyancy contour is shown in fig.\ref{fig:reamin_hfy_ra108}b. This term depends on the two-point temperature correlation and the direction of gravity. This term is present only in the y-direction heat flux density transport equation.}
\textcolor{black}{The Eqn \rf{eq:Era_rbc} and \rf{eq:J-rbc_simp} provide a mathematical means to analyze turbulent buoyancy-driven flows to deduce and quantify various scale space dynamics, hitherto inaccessible with single point statistics.    }




\section{Conclusion}
The Navier-Stokes (NS)  and energy equations describing the incompressible fluid flow give solutions in physical space. The velocity, pressure, and temperature fields computed using direct numerical simulations accommodate laminar, transition and turbulent solutions.
The turbulent flow field of such solutions contains mean quantities and fluctuating quantities, often resolved using Reynolds decomposition for incompressible flows.
The mean and fluctuating quantities, which are functions of space and time, however, do not give information about energy and heat transfer on the scale-space. This paper gives a mathematical description of the scale-space dynamics, which can be extracted from mean and fluctuating flow fields. 


Most turbulence theories and equations relating to statistics of the flow, like Kolmogorov's theories \cite{kolmogorov1941dissipation,frisch1995turbulence}, are asymptotic, valid for homogeneous isotropic turbulence. The energy transfer in scales in such flows can easily be computed using Fourier decomposition into energy contained in wavenumber space. Such a technique will not be possible for inhomogeneous anisotropic turbulent flow. Buoyancy-driven flows are inherently inhomogeneous, and we derived here the scale-space governing equation for such flows.

The derived transport equations are:\\
(1) two-point kinetic energy (scale energy) transport equation, \\
(2) two-point heat-flux (scale heat) transport equation,\\
(3) energy density transport equation, and 
\\(4) heat flux density transport equation.\\ These equations can be used to find out all the contributing factors to the balance of turbulence and, specifically, the contribution of local anisotropy and local inhomogeneity to energy transport or production. 

\textcolor{black}{The energy and heat-flux density transport equation can give us information not only on the amount of the energy/heat contained in different scales of turbulence at various locations in physical space but also on the scale and direction dependency of turbulence. They can provide detailed information on the energy/heat interaction between any two locations in the flow field.} These transport equations adequately describe the energy and heat flux in the scale space in a similar manner to the energy spectrum (energy cascade) in the wave number space. Hence, these equations extend the spectral energy density distribution concept to inhomogeneous and anisotropic flows. 

\textcolor{black}{The energy density and heat flux density transport equations are used to evaluate scale space energy transport in 2D RBC dataset. The production, scale space transport, dissipation and buoyancy source terms from the equations are evaluated for the RBC case of $Ra = 10^8$.}
The equations relate several statistics of turbulence hitherto unknown. Notably, the importance of inhomogeneous terms contributing to the interscale transport to higher and lower scales. \textcolor{black}{The equations relate the processes that change the energy at any eddy size or scale. The changes are due to the local production at that scale, the energy gain or loss from the processes that occur in physical space, and the energy transport processes in scale space.} The scale energy and scale heat flux dependence on scale size and orientation of the eddies is explicitly identified in the equations. The mathematical foundation used in the paper to ask questions related to scale transport is from fundamental principles. The equation provides a more generalized description of the physics of turbulence flow as a conceptual tool, augmenting the existing single-point statistical equation. These new terms in the scale transport equation will help in a more faithful modeling of turbulence. 


\section*{Acknowledgements}
The authors thank Ms. Esmitha Ganesan for proofreading and carefully checking all the mathematics for errors. We acknowledge Mr. K. K. Prasoon, who initiated this work.
\textcolor{black}{We thank Mr. K. Srikanth for the data set used in the case study.}

\section*{AUTHOR DECLARATIONS}
\subsection*{Conflict of Interest}
The authors have no conflicts to disclose.
\subsection*{Author Contributions}
\textbf{Endale H. Kirubel}: Methodology(equal); Writing-original draft(equal); Writing- review \& editing(equal). \textbf{P. Aswin}: Methodology(equal); Visualization(lead); Writing-original draft(equal); Writing - review \& editing(equal); Software(equal). \textbf{A. Sameen}: Supervision; Conceptualization(lead); Writing - review \& editing(equal). 

\section*{DATA AVAILABILITY}
The data that support the findings of this study are available from the corresponding author upon reasonable request.

\appendix
\textcolor{black}{\section{Energy density transport equation}\label{sec:appendixA}}
\textcolor{black}{To obtain the energy density transport equation, we apply $-\frac{1}{2} \pdr{}{r_{\alpha}}\lrs{Eqn~\rf{uu44_ii}}$ as,
\begin{align}
     &\frac{-1}{2}\pdr{}{r_\alpha}\lr{\pdr{Q_{ii}(\bm{x},\bm{r})}{t}} + \frac{-1}{2}\pdr{}{r_\alpha}\lr{\buk \x \pdr{Q_{ii}(\bm{x},\bm{r})}{x_k}} \nonumber \\
    &= {\frac{1}{2}\pdr{}{r_\alpha}\lr{ Q_{ki}(\bm{x},\bm{r})\pdr{\bui \x }{x_k}}} + {\frac{1}{2}\pdr{}{r_\alpha}\lr{\pdr{T_{iik}(\bm{x},\bm{r})}{x_k}}}\nonumber \\
    & - {\frac{1}{2}\pdr{}{r_\alpha}\lr{g_i \beta \lr{H_i(\bm{x},\bm{r}) + H_i'(\bm{x},\bm{r})}}}\nonumber \\
    & + {\frac{1}{2}\pdr{}{r_\alpha}\lr{\frac{1}{\rho}\ol{ \pdr{u_i'\xtt p'\x}{x_i}} }}   \nonumber \\
    & - {\frac{1}{2}\pdr{}{r_\alpha}\lr{\nu \pdr{}{x_k}\lr{\ol{u_i'\xtt \pdr{u_i'\x}{x_k}}}}}     \nonumber \\
    & + {\frac{1}{2}\pdr{}{r_\alpha}\lr{\nu \ol{\pdr{u_i'\xtt }{x_k}\pdr{u_i'\x}{x_k} }}}    \nonumber \\
    &+ {\frac{1}{2}\pdr{}{r_\alpha}\lr{Q_{ik}(\bm{x},\bm{r})\pdr{\bui \xtt}{r_k}}}    \nonumber \\
    & +{\frac{1}{2}\pdr{}{r_\alpha}\lr{\frac{1}{\rho}\ol{ u_i'\x\pdr{p'\xtt}{r_k}{  \delta{ik}}}}}     \nonumber \\
    & - {\frac{1}{2}\pdr{}{r_\alpha}\lr{\nu \pdr{}{r_k}\pdr{Q_{ii}(\bm{x},\bm{r}) }{r_k}}}     \nonumber \\
    &- {\frac{1}{2}\pdr{}{r_\alpha}\lr{\pdr{}{r_k}\lr{Q_{ii}(\bm{x},\bm{r})\lrs{\buk \x - \buk \xtt} }}}     \nonumber \\
    & - {\frac{1}{2}\pdr{}{r_\alpha}\lr{\pdr{}{r_k}\lr{\av{u_i'\x u_i'\xtt \lrs{u_k'\x - u_k'\xtt}}}}} ~.
     \label{uu45}
\end{align}
The heat flux density transport equation can be obtained in a similar way by taking the $r$ derivative of Eqn \rf{ut-rbc} as, 
\begin{align}
     &-\pdr{}{r_\alpha}\pdr{H_i}{t}   -\pdr{}{r_\alpha}\lr{\buk \x \pdr{H_i}{x_{k}}} =   \pdr{}{r_\alpha}\lr{\pdr{\mathcal{T}_{ik}}{x_k}}
     \nonumber \\ 
    &   + \pdr{}{r_\alpha}\lr{H_k\pdr{\bui\x}{x_{k}}}      +\pdr{}{r_\alpha}\lr{Q_{ik}\pdr{\ol{T}\xtt}{r_{k}}}
    \nonumber \\ 
    & -\pdr{}{r_\alpha}\lr{g_i\beta \ol{T'\xtt{T'\x}}} +  \frac{1}{\rho}\pdr{}{r_\alpha}\lr{\ol{ T'\xtt\pdr{ p'\x}{x_{k}}}\delta_{ik}}
    \nonumber  \\ 
     &  -  \nu \pdr{}{r_\alpha}{ \ol{ \pdr{}{x_k}\lr{T'\xtt \pdr{{u_i'\x}}{x_{k}}}}} +   \pdr{}{r_\alpha}\lr{ \ol{ \pdr{T'\xtt}{x_k}{ \pdr{{u_i'\x}}{x_{k}}}}} 
    \nonumber  \\ 
   & - \pdr{}{r_\alpha}\pdr{}{r_{k}}\lr{ \ol{u_i'\x  T'\xtt \lrs{u_k' \x -u_k' \xtt} }} 
   \nonumber \\
   & - \pdr{}{r_\alpha}\pdr{}{r_{k}}\lr{\lrs{\buk\x- \buk \xtt  }H_i} -   {\alpha \pdr{}{r_\alpha}{ \pdr{}{r_k}\lr{\pdr{H_i(x,r)}{r_{k}}}} }.
  \label{ut-rbc_2}
\end{align}
}

\textcolor{black}{\section{Two-point correlation transport equation for RBC flow}}
\textcolor{black}{Scale energy transport equation Eqn \rf{uu44_ii} and scale heat flux transport equation Eqn \rf{ut-rbc} in a dimensionless form for RBC with the scaling discussed in Sec \ref{sec:RBC} are written as,
\begin{align}
    & \pdr{Q_{ii}(x,r)}{t} +  \buk \x \pdr{Q_{ii}(x,r)}{x_k} = \underbrace{- Q_{ki}(x,r)\pdr{\bui \x }{x_k}}_{\bm{P}_x} 
    \nonumber \\
    & - \underbrace{ Q_{ik}(x,r)\pdr{\bui \xtt}{r_k}}_{\bm{P}_r} - \underbrace{\pdr{T_{iik}(x,r)}{x_k}}_{\bm{T}} + \underbrace{ PrEc \lr{H_i + H_i'} g_i}_{\bm{S}} 
    \nonumber \\
    &  - \underbrace{\frac{1}{Re}\ol{\pdr{u_i'\xtt }{x_k}\pdr{u_i'\x}{x_k} }}_{\bm{\epsilon}} + \underbrace{\frac{1}{Re}\pdr{}{x_k}\lr{\ol{u_i'\xtt \pdr{u_i'\x}{x_k}}}}_{\bm{D}^{\nu}_x} 
    \nonumber \\
    &  - \underbrace{\pdr{}{x_i}\lr{\ol{u_i'\xtt p'\x} }}_{\bm{D}^{p}_x} - \underbrace{\pdr{}{r_i}\lr{\av{u_i'\x p'\xtt }}}_{\bm{D}^p_r} 
    \nonumber \\
    & + \underbrace{\pdr{}{r_k}\lr{Q_{ii}\lrs{\buk \x - \buk \xtt} }}_{\bm{SST}_m} + \underbrace{\frac{1}{Re}\pdr{}{r_k}\pdr{Q_{ii}}{r_k}}_{\bm{\mathcal{D}}_r} 
    \nonumber \\
    &+ \underbrace{\pdr{}{r_k}\lr{\av{u_i'\x u_i'\xtt \lrs{u_k'\x - u_k'\xtt}}}}_{\bm{SST}_f} ~, 
    \label{Qii-rbc_nd}
\end{align}
and
\begin{align}
    &  \pdr{H_i}{t} + \buk \x \pdr{H_i}{x_{k}} = \underbrace{PrEc\ol{\theta'\xtt{\theta'\x}} g_i}_{\bm{S_\theta}} -\underbrace{H_k\pdr{\bui\x}{x_{k}}}_{\bm{P_{\theta x}}} 
    \nonumber \\
    & - \underbrace{Q_{ik}\pdr{\ol{\theta}\xtt}{r_{k}}}_{\bm{P_{\theta r}}} - \underbrace{\pdr{T_{ik}}{x_k}}_{\bm{T_\theta}} - \underbrace{\ol{ \theta'\xtt\pdr{ p'\x}{x_{k}}}\delta_{ik}}_{\bm{D^p_\theta}}
    \nonumber  \\ 
     &  + \underbrace{\frac{1}{Re} \ol{ \pdr{}{x_k}\lr{\theta'\xtt \pdr{{u_i'\x}}{x_{k}}}}}_{\bm{D^{\nu}_{\theta}}} 
     -  \underbrace{\frac{1}{Re} \ol{ \pdr{\theta'\xtt}{x_k}{ \pdr{{u_i'\x}}{x_{k}}}}}_{\bm{\epsilon_\theta}} 
    \nonumber  \\ +
   & \underbrace{ \pdr{}{r_{k}} \ol{u_i'\x  \theta'\xtt \lrs{u_k' \x -u_k' \xtt} }}_{\bm{SST_{\theta f}}}
   \nonumber \\ 
    &  + \underbrace{\pdr{}{r_{k}}\lrs{\buk\x- \buk \xtt  }H_i}_{\bm{SST_{\theta m}}} +  \underbrace{\frac{1}{\sqrt{PrRa}} { \pdr{}{r_k}\lr{\pdr{H_i(x,r)}{r_{k}}}} }_{\bm{\mathcal{D}_{\theta r}}} ~.
    \label{ut-rbc_nd}
\end{align}}

 \section*{References}

\end{document}